\documentclass[acmsmall,authorversion]{acmart} 

\usepackage{subcaption}
\usepackage{xcolor,colortbl}
\usepackage{multirow}
\usepackage{multicol}
\usepackage[nolist,nohyperlinks]{acronym}
\usepackage{lscape}

\usepackage[a-2b,mathxmp]{pdfx}[2018/12/22]

\newcommand{\bluecell}{\cellcolor[HTML]{CBCEFB}}
\newcommand{\redcell}{\cellcolor[HTML]{FFCCC9}}

\newacro{vir}  [VIR]  {view independent rendering}
\newacro{ivir}  [iVIR]  {improved view independent rendering}
\newacro{epr}  [EPR]  {eye-based point rendering}
\newacro{mvr}  [MVR]  {multiview rendering}
\newacro{pcss} [PCSS] {percentage-closer soft shadows}
\newacro{fps} [FPS] {frames per second}
\newacro{vr} [VR] {virtual reality}
\newacro{lfd} [LFD] {light field display}
\newacro{lfds} [LFDs] {light field displays}
\newacro{lfr} [LFR] {light field rendering}
\newacro{lfdpr} [LFDPR] {light field display point rendering}
\newacro{lfds} [LFDs] {light field displays}
\newacro{nlfds} [NLFDs] {Near-eye light field displays}
\newacro{gcd} [GCD] {gaze-contingent display}
\newacro{rmse} [RMSE] {root mean-squared error}
\newacro{eia} [EIA] {elemental image array}
\newacro{lod} [LOD] {level of detail}

\newcommand{\eref}[1]{Eq.~\ref{#1}}
\newcommand{\fref}[1]{Fig.~\ref{#1}}
\newcommand{\tref}[1]{Table~\ref{#1}}
\newcommand{\secref}[1]{Section~\ref{#1}}

\title{Light Field Display Point Rendering\thanks{This paper was published in ACM TOG 2024, DOI: 10.1145/3651300}}

\begin{document}

    \begin{teaserfigure}
\centering
    \includegraphics[width=1.0\linewidth]{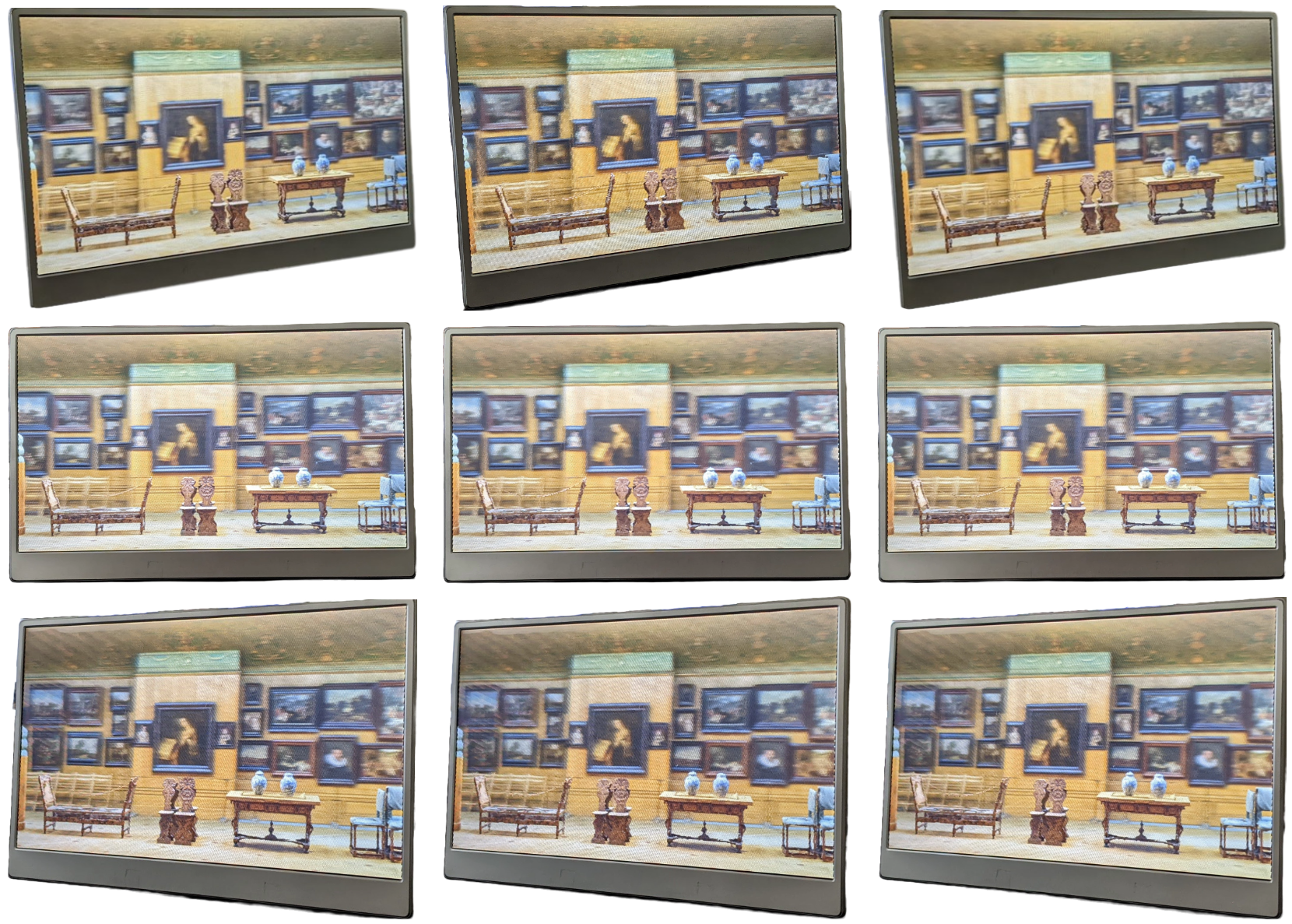}
  \caption{Results generated by our gold standard (GStd) (first column), light field display point rendering (LFDPR) (middle column), and standard multiview rendering (MVR) (last column)  as viewed through the lens from left (top row), center (middle row), and right (bottom row) viewing angles. Each renderer generates 48 distinct views. While they generate very similar imagery, LFDPR makes imagery up to 8$\times$ faster than MVR. }
  \label{fig:teaser}
\end{teaserfigure}

\author{Ajinkya Gavane}
\email{asgavane@ncsu.edu}
\orcid{0000-0002-3690-3899}
\affiliation{%
  \institution{North Carolina State University}
  \city{Raleigh}
  \country{United States}
}
\author{Benjamin Watson}
\email{bwatson@ncsu.edu}
\orcid{0000-0002-3758-7357}
\affiliation{%
  \institution{North Carolina State University}
  \city{Raleigh}
  \country{United States}
}

\renewcommand{\shortauthors}{Gavane and Watson}

    \begin{abstract}
Rendering for light field displays (LFDs) requires rendering of dozens or hundreds of views, which must then be combined into a single image on the display, making real-time LFD rendering extremely difficult. We introduce light field display point rendering (LFDPR), which meets these challenges by improving eye-based point rendering \cite{EPR2023} with texture-based splatting, which avoids oversampling of triangles mapped to only a few texels; and with LFD-biased sampling, which adjusts horizontal and vertical triangle sampling to match the sampling of the LFD itself. To improve image quality, we introduce multiview mipmapping, which reduces texture aliasing even though compute shaders do not support hardware mipmapping. We also introduce angular supersampling and reconstruction to combat LFD view aliasing and crosstalk. The resulting LFDPR is 2-8$\times$ times faster than multiview rendering, with similar comparable quality.
\end{abstract}




    \maketitle

        \section{Introduction}
    In emerging \ac{lfds}, each pixel emits several different colors, one for each of several angles of view, supporting unique views for stereoscopy and multiple simultaneous viewers, and simulating the experience of looking out (and around the edges) of a window. LFDs require multiview rendering to support angular variation in the light field they create.

    Graphics hardware is primarily optimized for rendering scenes from a single perspective, making multiview rendering challenging. \ac{vir}, \ac{ivir}, and \ac{epr}  \cite{marrs2017vir, iVIR2022, EPR2023}  avoid multiple passes by generating and rendering points in real time. Frame by frame, they convert the model's triangles into a set of points fit to the views the current frame needs. With the points, they then render views in parallel, requiring nearly an order of magnitude fewer passes.

    In this paper, we describe how we marshal and improve on the techniques of both \ac{ivir} and \ac{epr} to speed rendering for \ac{lfds} further. To achieve this, our \ac{lfdpr} includes the following innovations:
\begin{itemize}
    \item \textit{LFD-biased sampling} matches triangle sampling bias to LFD display sampling bias, reducing the size of the point cloud and improving efficiency.
    \item \textit{Texture-based splatting} samples triangles coarsely when the textures mapped to them are also coarse, again reducing the point cloud and improving rendering speed.
    \item \textit{Multiview mipmapping} supports texture antialiasing without requiring (currently unsupported) use of hardware mipmaps in the compute shader. This improves image quality with only a small impact on render speed. 
\end{itemize}
    
These innovations enable rendering of LFD views $2 - 8\times$ faster than standard \ac{mvr}, with nearly the same --- and sometimes better --- image quality.
        \section{Related Work}
\label{sec:relatedWork}

\begin{figure}[ht]
\centering
  \includegraphics[width=0.75\textwidth]{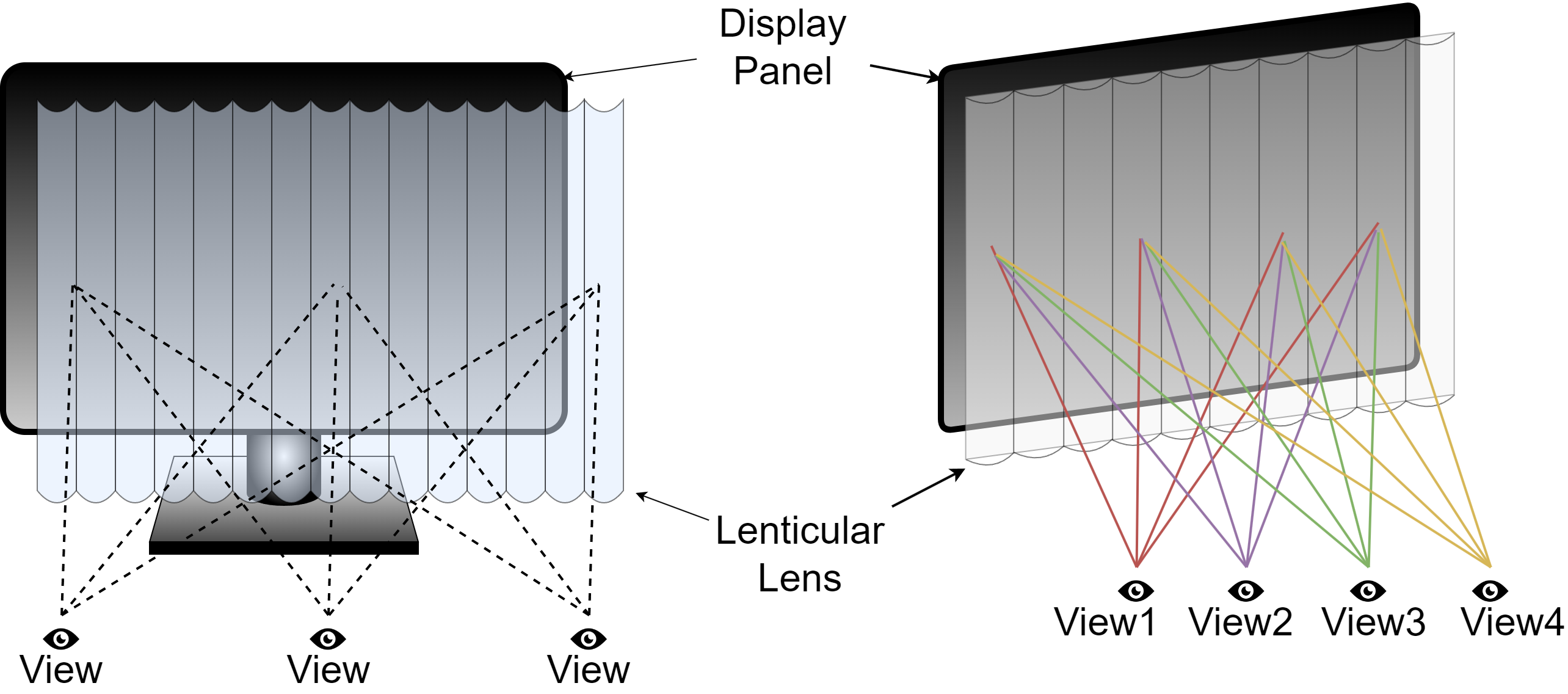}
  \caption{Viewpoints distributed across an LFD horizontally see different pixels and views through the lenticular lens. Each view corresponds to one panel pixel under each lenslet, creating a tradeoff between angular and spatial resolution.}
  \label{fig:autosteroscopicDisplays}
\end{figure}

Light field rendering aims to capture and reproduce all views of a scene, including not only how color varies across space, but also across angle.
At a given moment, the light field is radiance defined as a function of position and direction in an illuminated scene \cite{gershun1939lightfield}.  
Light field displays show the portion of the light field visible through the display surface, typically a rectangle. Light fields were first introduced to computer graphics in 1996 \cite{levoy1996lightfieldrendering,gortler1996lumigraphLightFieldRendering}. Since then, a great deal of research has followed, including generation of light fields with machine learning \cite{wang2018high, mildenhall2021nerf}, and capture of them with cameras \cite{wilburn2005cameraArrays, ng2005handheldLFD, marwah2013LFDCameraArchitecture}.

Our focus here is on \ac{lfds}. In most, each pixel represents not only the position and color of a light ray passing through the display surface, but also a range of orientations, with color varying across that range. Each pixel emits a different color based on the direction from which it is viewed, giving viewers different imagery as they move. Most often, to present views varying only horizontally, displays use parallax barriers that mask pixels not displaying the current view \cite{ives1903parallax}; or a lenticular lens, an array of narrow cylindrical lenses that redirects the light rays emitted by pixels to the same end \cite{matusik2004LFDOverview}. Both these types of LFDs offer stereoscopic and binocular depth (3D) cues, without requiring viewers to wear glasses or other technology. \fref{fig:autosteroscopicDisplays} demonstrates light flow for lenticular lens display systems. Note that such support for multiple views on a single display panel introduces a tradeoff between spatial and angular resolution.

\ac{lfds} supporting horizontal view change are now commercially available. Examples include products by FOVi3D, Dimenco and LeiaInc \cite{DimencoWebsite}, Looking Glass Display \cite{LookingGlassFactory2023}, Light Field Lab \cite{lightFieldLabs2018} and AYE3D \cite{Aye3D_2022}.
In particular, LeiaInc and Dimenco recently collaborated with ASUS and launched 16" \ac{lfd} laptops. \ac{lfds} are also finding their way into head-mounted displays (HMDs). NVIDIA proposed a near-eye \ac{lfd} \cite{lanman2013neareyelfd} that is light-weight and slim, incorporated into eyeglasses. These displays use the angular flexibility of light fields to eliminate the need for heavy, fixed lenses; to allow the use of HMDs without prescription eyeglasses; and to mitigate the vergence-accommodation conflict faced in most HMDs. 

\begin{figure}[t]
    \centering
    \includegraphics[width=0.95\textwidth]{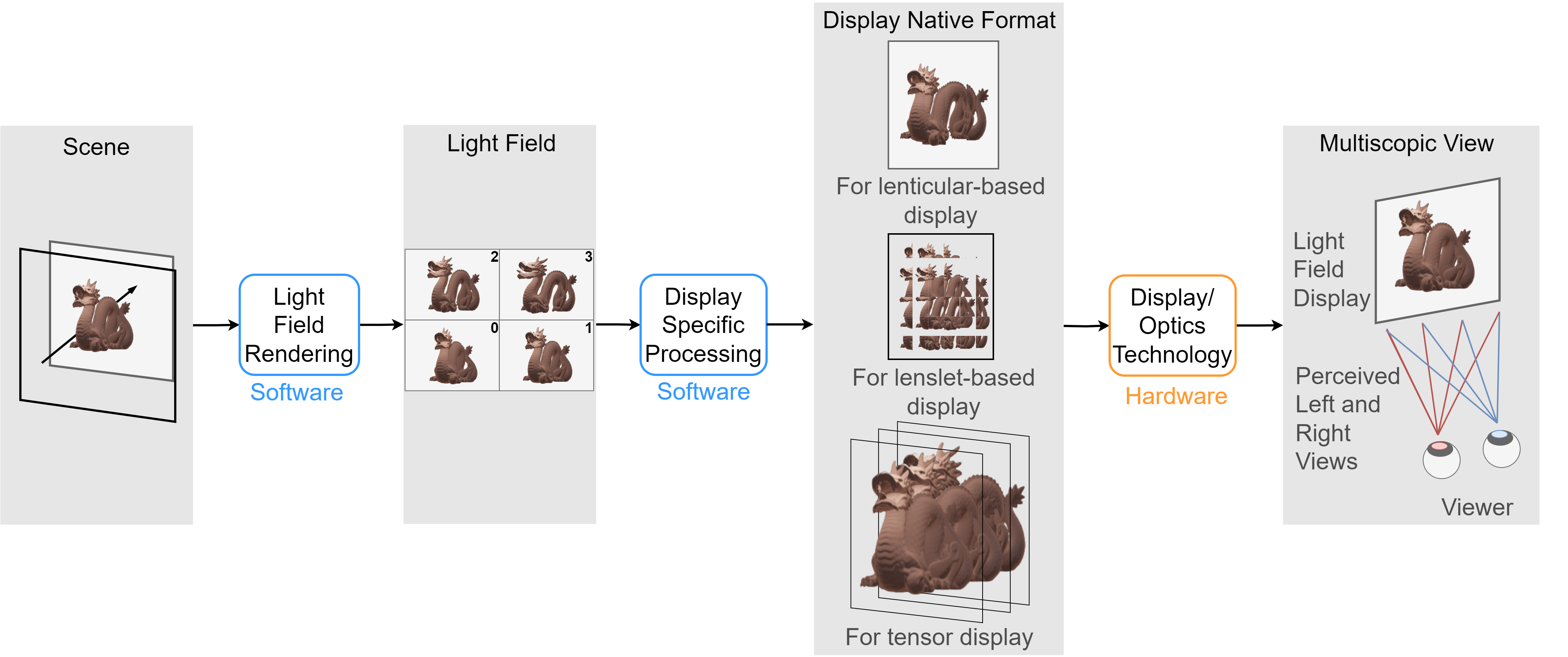}
    \caption{Conventional image formation process for light field displays. A 3D scene is processed to generate a fronto-parallel light field which is then processed to be compatible with the \ac{lfd}. When viewed from the \ac{lfd}, the displayed image is perceived as a 3D scene \cite{fink2023_i3dLFD}.}
    \label{fig:RW_StdLFDRenderPipeline}
\end{figure}

\subsection{Light Field Rendering}

Light field rendering is the process of generating synthetic light fields that form input for LFDs. \fref{fig:RW_StdLFDRenderPipeline}  illustrates the full LFD image formation process; light field rendering is the first step it depicts.

To produce visual content for \ac{lfds}, it is crucial to capture the 3D spatial data of a scene from various viewing angles. This can be achieved by configuring a virtual camera array to capture a scene from multiple viewing directions. The resulting views are then combined into a format compatible with a native \ac{lfd} presentation \cite{fink2023_i3dLFD}. One of the first examples is Levoy et al. \cite{levoy1996lightfieldrendering}, who simplified the 7D plenoptic function \cite{adelson1991plenoptic} (six degrees of angular and spatial freedom and radiance), into a 4D representation (a matrix of 2D images). This representation allowed computational efficiencies, uniform sampling of light fields, and control over the set of rays.

The standard approach to rendering light fields with GPUs is to make multiple passes over the scene description, with one pass per view (standard \ac{mvr}) \cite{unterguggenberger2020fastMVR_RT}. Because this is quite slow, H{\"u}bner et al. \cite{hubner2006MVRpointsplatting}  proposed a single-pass solution that generates multiview splats and performs per-pixel ray-splat intersections in the fragment shader. In \cite{hubner2007singlepassMVVR}, they improved performance further with volume rendering method based on 3D textures. But their approach cannot directly render triangles and produces a maximum of eight views --- not enough for modern LFDs. Sorbier et al. \cite{de2010GPUbasedMVR} duplicated geometry to reduce CPU-GPU bandwidth, rendering twice as fast as standard \ac{mvr} for 80k triangles and $20$ views. Our work also uses points to accelerate light field rendering with GPUs, but achieves much higher speedups.

Recently, AI has been employed to rapidly generate high-quality light fields within a specific viewing range for static scenes \cite{Flynn2019DeepViewSynthesisLFDAI, mildenhall2021nerf,Wizadwongsa2021NexLFDAI,wang2018high} or even for scenes that have views changing over time \cite{Bemana2020X-Fields_LFDAI, suhail2022LFNeuralRend_LFDAI}. However, these approaches have limitations: they do not support scenes with moving objects, and were applied to scenes captured with cameras, rather than scenes rendered in real time. There is still considerable progress needed to achieve real-time rendering of high-quality content with AI-generated light fields.

\subsection{Points, \ac{ivir}, and \ac{epr}}
Using points as a rendering primitive can avoid many of the limitations of triangle rasterization \cite{levoy1985points,gross2011pointbasedgraphics}. 
Unfortunately when views change, surface gaps can appear in projected point clouds. Preventing such artifacts often requires a prohibitive number of points, or similarly prohibitive filtering of sparser point sets (e.g., \cite{ritschel2008imperfectShadowMap}). However, newer algorithms can avoid such painful tradeoffs. Ritschel et al. \cite{ritschel2008imperfectShadowMap,ritschel2009micro,ritschel2011improved_ISM} adaptively sample according to brightness and view. Schutz et al.  \cite{schutz2021rendering} apply compute shaders to render points more quickly.

Marrs et al.'s \ac{vir} \cite{marrs2017vir} utilized the GPU rasterizer to generate frame-specific point clouds for off-screen views, enabling parallel rendering of views with significantly fewer geometry traversals than MVR. However, as views become more heterogeneous (as with environment maps or omnidirectional shadows), point cloud size grows, making real-time performance challenging. Gavane and Watson's \ac{ivir} \cite{iVIR2022} addresses this by optimizing triangle sampling and resizing off-screen buffers to match eye buffer resolution. This reduces the number of points and shader loads, enabling rendering of environment maps $2-4\times$ faster than MVR.

Gavane and Watson's \ac{epr} \cite{EPR2023} further improved the performance and generality of point rendering for multiview effects. \ac{epr}  tailors point clouds to the sampling rate required by the eye (frame) buffer instead of off-screen buffers, and by splatting resulting points across many off-screen pixels. \ac{epr} also deferred shading further by calculating off-screen lighting only when visible to the eye, and enabled very fast no-pass recursive reflection for dynamic environment mapping. As a result, \ac{epr} is $6-7\times$ faster than MVR and iVIR, with nearly identical image quality.

        \section{LFD Point Rendering}
\label{sec:lfdpr}
\begin{figure}[t]
    \centering
    \includegraphics[width=1.0\textwidth]{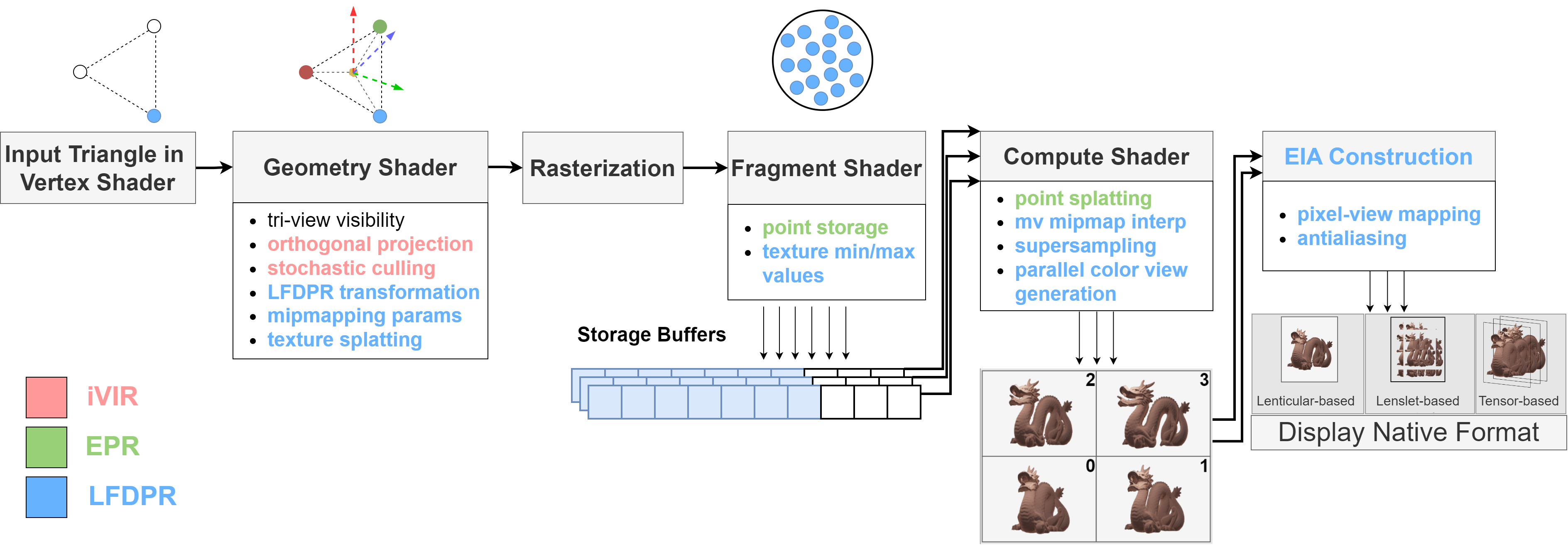}
    \caption{LFDPR pipeline as implemented in the GPU. Improvements of iVIR and EPR are marked in red and green font respectively, and LFDPR innovations are marked in blue font.}
    \label{fig:LFDPRpipeline}
\end{figure}

\ac{lfdpr} combines \ac{ivir}'s triangle sampling with \ac{epr}'s point splatting to form a foundation for rendering \ac{lfd} views.
\fref{fig:LFDPRpipeline} shows an overview of the LFDPR pipeline. Like \ac{ivir} \cite{iVIR2022}, the vertex shader checks vertex visibility, but uses views defined by the LFD rather than shadows or environment maps. Also like \ac{ivir}, the geometry shader adjusts sampling density, but to LFD view images with rectangular aspect ratios, and to match texture density (\secref{section:lfd_samplingDensity}). 
Similar to \ac{epr}, LFDPR's fragment shader often splats points across multiple pixels when textures are coarser than the view buffer. However, it also implements multiview mipmapping (\secref{section:lfd_mvMipmapping}). Finally, a single image pass constructs the \ac{eia}, which interweaves rendered images for display on the panel, and viewing through the LFD's lenticular lens.

\subsection{Determining Sampling Density}
\label{section:lfd_samplingDensity}

The geometry shader adjusts the point sampling density for each triangle using iVIR logic \cite{iVIR2022} by computing the projection matrix $T_{align}$ applied to the triangle. However, LFDPR introduces three improvements: sampling according to LFD aspect ratio, texture-based sampling, and setup for multiview mipmapping.

With the angular-spatial tradeoff LFDs introduce (see \secref{sec:relatedWork}), the individual views generated by LFD rendering often have rectangular aspect ratios, with more pixels in the vertical dimension. Furthermore, unlike off-screen views for shadows and reflections, the centers of projection for these views change predictably, being distributed horizontally across the LFD. Thus, triangles need not be sampled as densely in the horizontal as in the vertical direction.

\begin{figure}[t]
  \centering
  \includegraphics[width=0.6\linewidth]{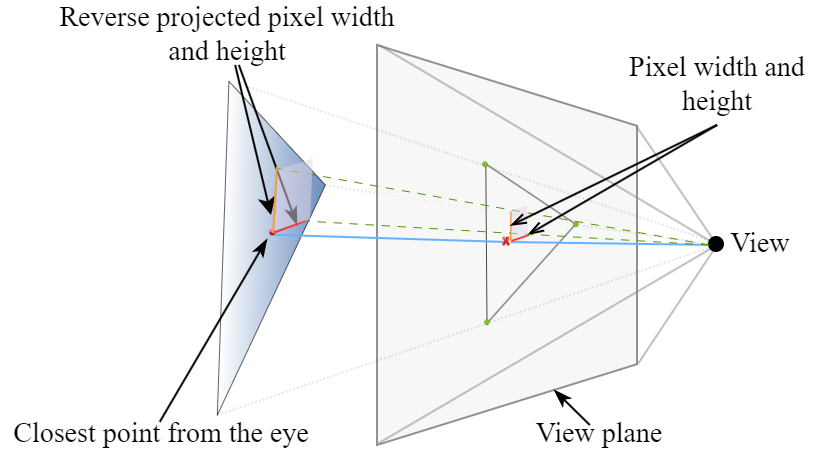}
  \caption{Reverse projection of pixel edges around the closest point on the polygon from the eye. The horizontal and vertical pixel edges often differ in length on LFDs.}
   \label{fig:LFD2Dsampling}
\end{figure}

To leverage this fact, LFDPR not only aligns triangles parallel to \ac{ivir}'s view point-sampling plane, it also aligns the vertical and horizontal axes in iVIR's sampling view to the vertical and horizontal axes in the light field views. Thus $T_{align}$ \cite{iVIR2022} not only orients the triangle's normal perpendicular to \ac{ivir}'s view plane, but also maintains two sampling densities ($\vec{S}{_{ortho}}$) for the horizontal and vertical dimensions. 
To achieve this, instead of computing sampling density per unit area, we compute it horizontally and vertically by projecting pixel edges, see \fref{fig:LFD2Dsampling}.

To make LFDPR sampling still more parsimonious, we introduce texture-based sampling. In \ac{ivir} and \ac{epr}, many points required to meet buffer sampling requirements may in fact be sampling the same texels, resulting in the same color. Texture-based sampling reduces point density to avoid this.
In the geometry shader, we compute the maximum texel density on the triangle (denoted as $size_{uv}$) by dividing the area of triangle in the world space, $A_{ws}$,  by the maximum number of texels $n_{tex}$ a triangle spans over all the textures attached to it. This value is independent of any particular LFD view. \eref{eq:texelCount} and \eref{eq:lfd_UVSize} describe this process.
    
\begin{equation}
    n_{tex} = \max_{t \in T}\left( A_{ts} \times X_t \times Y_t \right)
\label{eq:texelCount}
\end{equation}
\begin{equation}
    size_{uv} = \left(\frac{A_{ws}}{n_{tex}}\right)
\label{eq:lfd_UVSize}
\end{equation}

\noindent where $A_{ts}$ is the area of triangle in texture space, and T is the set  of all the textures   attached to a triangle with each texture t of resolution $[X_t, Y_t]$. In iVIR, triangles are scaled by ${S}{_{ortho}}$ to set the point sampling rate \cite{iVIR2022}. In LFDPR, when the $size_{uv}$ is greater than the size of the reverse projected view pixel $size_{xy}$, we adjust our sampling density using the formula given in Equation \eqref{eq:lfd_texelSplatDensity}.

\begin{equation}    
 \vec{S}{_{ortho}} = 
\begin{cases}
    \vec{S}{_{ortho}} \times \sqrt{\frac{size_{xy}}{size_{uv}}} & \text{if } size_{uv} > size_{xy} \\
    \vec{S}{_{ortho}} & \text{otherwise}
\end{cases}
\label{eq:lfd_texelSplatDensity}
\end{equation}

The geometry shader also sets up multiview mipmapping, discussed in \secref{section:lfd_mvMipmapping}.

%
%
\subsection{Splatting}
\label{section:lfd_splatting}

With texture-based sampling producing points sized larger than buffer pixels, we can no longer rely on iVIR's simple point projection. Instead, LFDPR uses EPR's splatting logic in the compute shader \cite{EPR2023}. To splat, the shader projects a world-space square into the current view, sized according to splat length and centered at the point, resulting in a 2D quadrilateral. To clip the quadrilateral against the triangle that contains it, the computer shader fills only off-screen buffer pixels lying inside the triangle.
This not only avoids holes between points, but also improves image quality by enabling use of EPR's inside-triangle test during point splatting \cite{EPR2023}, resulting in near-rasterization edge quality. This is particularly important for light field rendering, when view and display are sampled at nearly the same rate, in contrast to shadow and reflection buffers.

\subsection{Multiview Mipmapping}
\label{section:lfd_mvMipmapping}
\begin{figure}[t]
    \centering
    \includegraphics[width=1.0\textwidth]{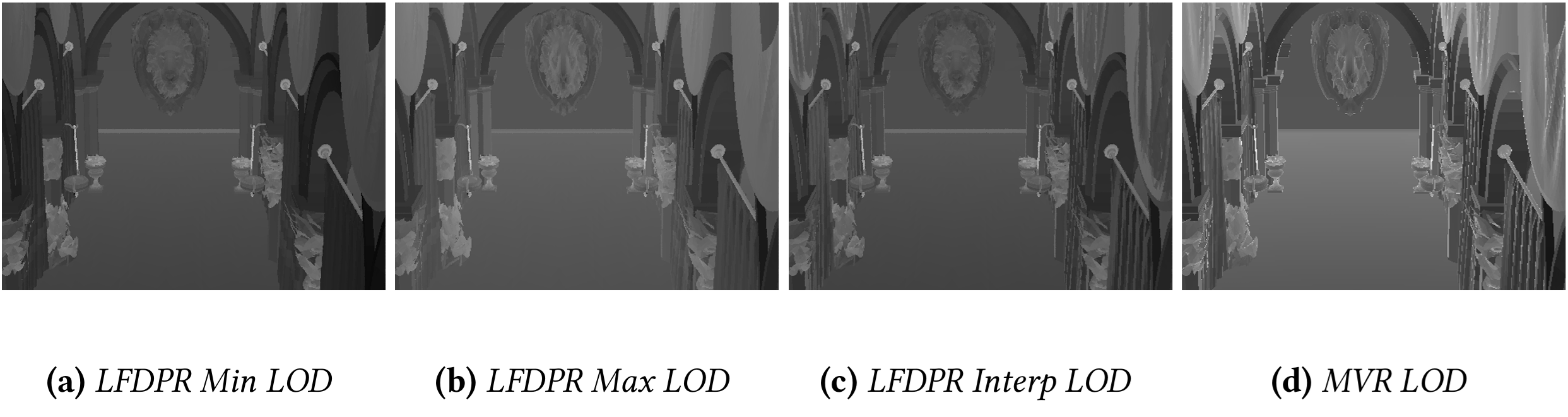}
\caption{The proportion of an entire texture covered by pixels (LOD maps), as generated by \ac{lfdpr} and \ac{mvr}. \ac{lfdpr}'s multiview mipmapping generates two \ac{lod} values for each texture in each triangle: (a) the minimum \ac{lod} over all views, and (b) the maximum \ac{lod} over all views. In (c), a particular view's LOD map is produced by interpolating between the minimum and maximum \ac{lod} values. (d) presents the corresponding \ac{lod} map in \ac{mvr}.}
\label{fig:LFD_lodMaps}
\end{figure}

When texturing, mipmapping ensures good image quality by removing much of the aliasing that results when many texels project to the same view pixel. In iVIR and EPR, multiview point splatting in the compute shader does not support hardware mipmapping, but only shadows and reflections were rendered with points, leaving the majority of the eye view mipmapped. However when rendering light fields, every visible pixel is rendered with points, making the lack of mipmapping much more problematic. We introduce multiview mipmapping to address this problem.

When the geometry shader is processing a triangle, we compute both the minimum and maximum size any pixel on the triangle reaches in texture space across all views, represented as $size_{vpuv}^{min}$ and $size_{vpuv}^{max}$ respectively\footnote{When triangle textures do not share the same mapping, $size_{vpuv}^{min}$ and $size_{vpuv}^{max}$ will also depend on texture.}$^,$\footnote{These equations assume that geometric and texture space are proportionally isomorphic across the triangle, e.g. we assume that when X changes by 10\%, so does U.}.

\begin{equation}
    size_{vpuv}^{min} = \min_{v \in V}\left(size_{xy}^v \times \frac{A_{ts}}{A_{ws}}\right)
    \label{eq:minVPUVSize}
\end{equation}
\begin{equation}
    size_{vpuv}^{max} = \max_{v \in V}\left(size_{xy}^v \times \frac{A_{ts}}{A_{ws}}\right)
    \label{eq:maxVPUVSize}
\end{equation}
\noindent where $size_{xy}^v$ is the area of a pixel from view $v$ projected into the triangle.

\eref{eq:minLOD} and \eref{eq:maxLOD} show how the fragment shader transforms these size variables into the range typically used by graphics libraries such as OpenGL, often called level of detail or \ac{lod}. Given that each texture may have varying resolutions (e.g., albedo, normal, roughness, ao, metallic, etc.), the fragment shader calculates the minimum ($LOD_{min}^t$) and maximum ($LOD_{max}^t$) LOD values for each texture, uses those values to fetch minimum and maximum texture values from each texture's hardware mipmap ($tv_{min}$ and $tv_{max}$), and stores those in the point buffer along with $size_{vpuv}^{min}$ and $size_{vpuv}^{max}$.  Fig. 6a and Fig. 6b show min and max \ac{lod} maps for the sponza scene. 

\begin{equation}
    LOD_{min}^t = 0.5 \times log_2\left(size_{vpuv}^{min}\times X_t \times Y_t \right)
    \label{eq:minLOD}
\end{equation}
\begin{equation}
    LOD_{max}^t = 0.5 \times log_2\left(size_{vpuv}^{max}\times X_t \times Y_t \right)
    \label{eq:maxLOD}
\end{equation}

During view generation in the compute shader, for each combination of view and point splat we interpolate a distinct displayed texture value \textit{tv} in the range [$tv_{min}$,$tv_{max}$], as described in \eref{eq:interplateTexValue}. To perform this interpolation, we must first estimate the amount of texture the point splat covers in the current view, $size^{vs}_{uv}$. 
To do this, we take the area of the splat in texture space, and increase it in inverse proportion to the part of the view pixel the splat covers (we never decrease a splat's texture area, since we render per splat, not per pixel). We then clamp the area to the range [$size_{vpuv}^{min}$, $size_{vpuv}^{max}$]. In \eref{eq:interpolationParameter}, $size^{vse}_{uv}$ is the estimated texture area adjusted by the proportion of the pixel it covers, $size^{vs}_{ndc}$ is the size of the splat in view space, in normalized device coordinates, and $X_v$ and $Y_v$ are the view resolution. 

\begin{figure}[t]
\centering
    \includegraphics[width=1.0\textwidth]{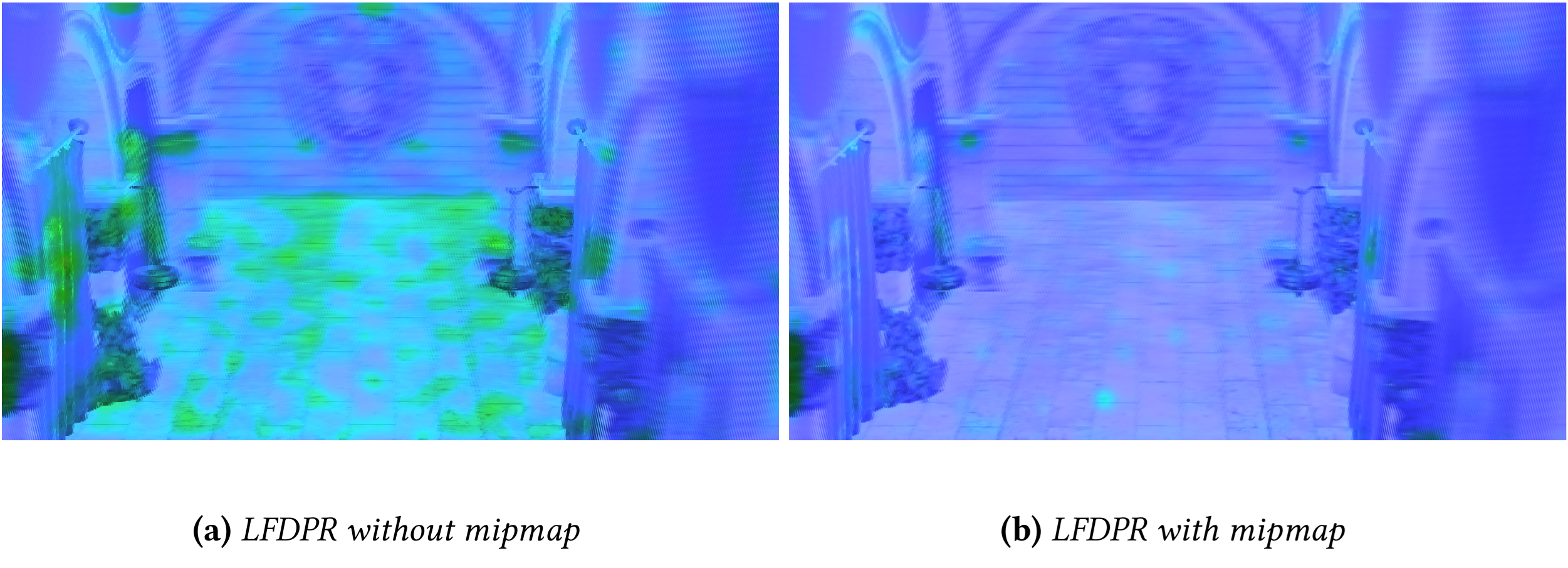}
\caption{Perceptual quality of the EIA rendered by LFDPR (a) without multiview mipmapping and (b) with it. Here, HDR-VDP3 compares each image against our MVR gold standard, with blue indicating rarely perceived differences, green slightly more perceivable, and red usually seen.}
\label{fig:LFD_mipmapResult}
\end{figure}

\begin{align}
\begin{split}
    size^{vse}_{uv} &=  \frac{size_{uv}}{\min(1,  size^{vs}_{ndc} \times 0.25 \times X_v  \times Y_v )}  \\
    size^{vs}_{uv} &= \text{clamp}(size^{vse}_{uv}, size_{vpuv}^{min}, size_{vpuv}^{max})
\end{split}
    \label{eq:interpolationParameter}
\end{align}

\begin{equation}
    tv = tv_{min} + \left( tv_{max} - tv_{min} \right) \times \left(\frac{size^{vs}_{uv} - size_{vpuv}^{min}}{size_{vpuv}^{max} - size_{vpuv}^{min}} \right)
    \label{eq:interplateTexValue}
\end{equation}

Based on the above interpolated LOD value, we shade each view in the compute shader. In contrast to EPR \cite{EPR2023}, we found that shading during rendering of views was more efficient than deferring shading to the final interleaving pass (see \secref{section:interleaving}), likely because texture access in the EIA was more incoherent. 
 Fig. 6c shows the interpolation of the \ac{lod} in a view, and \fref{fig:LFD_mipmapResult} shows the HDR-VDP3 perceptual quality map \cite{mantiuk2023hdrvdp3} comparing an LFDPR rendering of an EIA for the sponza scene with and without our mipmapping against the \ac{mvr} gold standard (GStd) --- a slowly rendered, very high quality image we use to evaluate image quality.

\section{Interleaving, reconstruction, and supersampling}
\label{section:interleaving}
Having rendered the necessary view images with LFDPR, we construct the EIA for the LFD by interleaving the views in a final image pass, as described in \cite{fink2023_i3dLFD}. As we built these EIAs, we sometimes applied antialiasing and supersampling \cite{moller2018realTimeRendering, crow1977aliasing}. While the benefits of these techniques in space are familiar, the value of their angular analogs are less well known.
For reconstruction, we used Gaussian filters with a spatial diameter of 2.25 pixels, and an angular support of 2 view intervals. To integrate these filters, we sampled them 8 times using a random Gaussian distribution. We also investigated the value of supersampling, including modest $2\times$ supersampling across space (view images with twice the resolution) and across angle (with splats projected into twice as many views, at random angles in view intervals). 
We discuss the benefits of this reconstruction and supersampling in \secref{sec:results}.
        
\section{Results}
\label{sec:results}
In this section, we assess the effectiveness of \ac{lfdpr} as measured by memory usage, rendering quality, and computation time. We begin by outlining the experimental setup we employed. Subsequently, we delve into rendering performance and image quality.

\subsection{Experimental Setup}
\label{sec:ch5ExpSetup}
We compared the light fields made by \ac{lfdpr} and MVR. We used OpenGL 4.5 on a PC with an Intel i5-7600K @ 3.80 GHz CPU and an NVIDIA RTX 3070 GPU, running Windows 10. We rendered the sponza, gallery, car, and coconut scenes \cite{McGuire2017Data, coconutScene} shown in \fref{fig:LFD_1xSS}. 
The sponza and gallery scenes are like many used in games, filling the entire view and making heavy use of textures. The gallery scene in particular has very large (16K$\times$16K) textures. Additionally, we tessellated large triangles in these scenes, to ensure that they spanned no more than a hundred pixels. As in both \ac{ivir} and \ac{epr}, traditional rasterization is more efficient when triangles are so large. The coconuts do not fill the entire view, but still have a large number of triangles, and large textures. Finally the car has fewer triangles and no textures, but significant depth complexity (occluded components). All the scenes were dynamic, rotating around their centers, and used a physically-based rendering shader \cite{pharr2016PBR_BOOK} that accessed albedo, roughness, metallic and normal textures. 

Our light field views used $64$-bit unsigned integer buffers, with a resolution of $480\times360$ (and $960\times720$ for $2\times$ spatial supersampling). The \ac{mvr} implementation rendered the light field views using multiple pipeline passes, each using hardware mipmapping with the same resolution of $480\times360$ (and $960\times720$ for $2\times$ spatial supersampling). Unlike \cite{unterguggenberger2020fastMVR_RT}, we did not optimize our \ac{mvr} implementation. Additionally, to support supersampling we built a mipmap of each light field view.  We compare all of our results to a slowly rendered, high quality \ac{mvr} implementation with 96 4K resolution views using anisotropic mipmapping and Gaussian spatial reconstruction filters, sampled 32 times using a random Gaussian sampling scheme. We call this high quality image our gold standard (GStd).

We displayed our rendering results on a custom-built LFD prototype with a tilted lens  \cite{yu2014-DenseViews_tiltedvsNon}, with a tilt angle of -9.66 degrees, 3840 x 2160 resolution, screen width of 345.40 mm, height of 194.30 mm, lens width of 0.72 mm, pixel pitch of 0.09 mm, subpixel pitch of 0.03 mm, and lens count (not an actual count, rather the ratio of the screen and lens widths) of 479.36. Note that because of imperfections in manufacture, this value differs slightly from the ideal lens count of 479.72. Similar imperfections occur even in professionally manufactured LFDs such as those made by Looking Glass.

\subsection{Memory Comparison}

In \ac{lfdpr}, each view buffer required $1.32$MB of memory; \ac{mvr}'s needed $2.63$MB. \ac{lfdpr} required $63.36$MB for $48$ view buffers and $80$MB per million points; \ac{mvr} required $126.24$MB for $48$ view buffers. Current GPU memories can contain more than a dozen GB.

\subsection{Speed and Error Comparison}

To compare performance, we averaged GPU run-time and the number of points generated over 1000 frames, with each technique generating the same views.
Tables \ref{table:lfd_1xResults} and \ref{table:lfd_2xResults} compare \ac{lfdpr}'s speed to \ac{mvr}'s for the sponza, gallery, coconut, and car scene with no and $2\times$ supersampling, respectively. 


\begin{table}[t]
\centering
\resizebox{\linewidth}{!}{
 \begin{tabular}{ c | c | c  c  c  c | c  c  c  c  c|}
 \hline
     \multicolumn{1}{c|}{\multirow{3}{*}{\shortstack[s]{Scene \\ (\#tris)}}} 
   & \multicolumn{1}{c|}{\multirow{3}{*}{\shortstack[s]{ Method }}} 
   & \multicolumn{4}{c|}{ Performance (in milliseconds)    }
   & \multicolumn{5}{c|}{ Error / Quality for EIA (Single View) }
   \\
   \cline{3-11} 

    & 
    & \begin{tabular}[c]{@{}c@{}}  pt gen \\ \textsuperscript{(\#pts)} \end{tabular}
    & \begin{tabular}[c]{@{}c@{}} view gen \end{tabular}
    & \begin{tabular}[c]{@{}c@{}} EIA \\ constr \end{tabular}
    & \begin{tabular}[c]{@{}c@{}}  tot \\ time  \end{tabular}
    & \begin{tabular}[c]{@{}c@{}}  RMSE  \end{tabular}
    & \begin{tabular}[c]{@{}c@{}}  P-det  \end{tabular}
    & \begin{tabular}[c]{@{}c@{}}  Q  \end{tabular}
    & \begin{tabular}[c]{@{}c@{}}  Q-JOD  \end{tabular}
    & \begin{tabular}[c]{@{}c@{}}  SSIM  \end{tabular}
    \\ 
    \hline \hline

    \multicolumn{1}{c|}{\multirow{10}{*}{\shortstack[s]{ Sponza \\ (1.0M) }}} 
    & \begin{tabular}[c]{@{}c@{}} MVR \end{tabular}
    & \begin{tabular}[c]{@{}c@{}} -  \end{tabular}
    & \begin{tabular}[c]{@{}c@{}} 30.74 \end{tabular}
    & \begin{tabular}[c]{@{}c@{}} 8.18 \end{tabular}
    & \redcell \begin{tabular}[c]{@{}c@{}} 38.92  \end{tabular}
    & \begin{tabular}[c]{@{}c@{}} 1.81 \\ (1.52)  \end{tabular}
    & \begin{tabular}[c]{@{}c@{}} 0.74 \\ (0.73)  \end{tabular}
    & \begin{tabular}[c]{@{}c@{}} 9.84 \\ (8.83) \end{tabular}
    & \begin{tabular}[c]{@{}c@{}} 9.95 \\ (9.36) \end{tabular}
    & \begin{tabular}[c]{@{}c@{}} 0.91 \\ (0.93)  \end{tabular}
    \\
    \cline{2-11}
    & \begin{tabular}[c]{@{}c@{}} LFDPR \\ no AA \end{tabular}
    & \begin{tabular}[c]{@{}c@{}} $\underset{(2.29M)}{2.81}$ \end{tabular}
    & \begin{tabular}[c]{@{}c@{}} 9.57 \end{tabular}
    & \begin{tabular}[c]{@{}c@{}} 0.43 \end{tabular}
    & \bluecell \begin{tabular}[c]{@{}c@{}} $\underset{(3.04\times)}{12.81}$  \end{tabular} 
    & \begin{tabular}[c]{@{}c@{}} 2.30 \\ (1.91)  \end{tabular}
    & \begin{tabular}[c]{@{}c@{}} 0.34 \\ (0.70)  \end{tabular}
    & \begin{tabular}[c]{@{}c@{}} 9.66 \\ (8.56) \end{tabular}
    & \begin{tabular}[c]{@{}c@{}} 9.87 \\ (9.17) \end{tabular}
    & \begin{tabular}[c]{@{}c@{}} 0.86 \\ (0.90)  \end{tabular}
    \\
    \cline{2-11}
    & \begin{tabular}[c]{@{}c@{}}  LFDPR  \\ Sp AA \end{tabular}
    & \begin{tabular}[c]{@{}c@{}} $\underset{(2.29M)}{2.78}$ \end{tabular}
    & \begin{tabular}[c]{@{}c@{}} 9.70 \end{tabular}
    & \begin{tabular}[c]{@{}c@{}} 1.51 \end{tabular}
    & \bluecell \begin{tabular}[c]{@{}c@{}} $\underset{(2.78\times)}{13.99}$  \end{tabular}
    & \begin{tabular}[c]{@{}c@{}} 2.07 \\ (1.86)  \end{tabular}
    & \begin{tabular}[c]{@{}c@{}} 0.32 \\ (0.68)  \end{tabular}
    & \begin{tabular}[c]{@{}c@{}} 9.75 \\ (8.60) \end{tabular}
    & \begin{tabular}[c]{@{}c@{}} 9.91 \\ (9.20) \end{tabular}
    & \begin{tabular}[c]{@{}c@{}} 0.89 \\ (0.91)  \end{tabular}
    \\
    \cline{2-11}
    & \begin{tabular}[c]{@{}c@{}}  LFDPR \\ View AA \end{tabular}
    & \begin{tabular}[c]{@{}c@{}} $\underset{(2.29M)}{2.72}$ \end{tabular}
    & \begin{tabular}[c]{@{}c@{}} 9.69 \end{tabular}
    & \begin{tabular}[c]{@{}c@{}} 1.27 \end{tabular}
    & \bluecell \begin{tabular}[c]{@{}c@{}} $\underset{(2.84\times)}{13.68}$   \end{tabular}
    & \begin{tabular}[c]{@{}c@{}} 2.29 \\ (1.91)  \end{tabular}
    & \begin{tabular}[c]{@{}c@{}} 0.34 \\ (0.71)  \end{tabular}
    & \begin{tabular}[c]{@{}c@{}} 9.65 \\ (8.56) \end{tabular}
    & \begin{tabular}[c]{@{}c@{}} 9.87 \\ (9.17) \end{tabular}
    & \begin{tabular}[c]{@{}c@{}} 0.87 \\ (0.90)  \end{tabular}
    \\
    \cline{2-11}
    & \begin{tabular}[c]{@{}c@{}}  LFDPR  \\ Vi-Sp AA \end{tabular}
    & \begin{tabular}[c]{@{}c@{}} $\underset{(2.29M)}{2.73}$\end{tabular}
    & \begin{tabular}[c]{@{}c@{}} 9.71 \end{tabular}
    & \begin{tabular}[c]{@{}c@{}} 2.16 \end{tabular}
    & \bluecell \begin{tabular}[c]{@{}c@{}} $\underset{(2.66\times)}{14.60}$   \end{tabular}
    & \begin{tabular}[c]{@{}c@{}} 2.07 \\ (1.87)  \end{tabular}
    & \begin{tabular}[c]{@{}c@{}} 0.32 \\ (0.67)  \end{tabular}
    & \begin{tabular}[c]{@{}c@{}} 9.75 \\ (8.59) \end{tabular}
    & \begin{tabular}[c]{@{}c@{}} 9.91 \\ (9.19) \end{tabular}
    & \begin{tabular}[c]{@{}c@{}} 0.88 \\ (0.90)  \end{tabular}
    \\
    \hline 
    \hline

    \multicolumn{1}{c|}{\multirow{9}{*}{\shortstack[s]{ Gallery \\ (998.9K) }}} 
    & \begin{tabular}[c]{@{}c@{}}  MVR \end{tabular}
    & \begin{tabular}[c]{@{}c@{}} -  \end{tabular}
    & \begin{tabular}[c]{@{}c@{}} 25.06 \end{tabular}
    & \begin{tabular}[c]{@{}c@{}} 8.20 \end{tabular}
    & \redcell \begin{tabular}[c]{@{}c@{}} 33.26  \end{tabular}
    & \begin{tabular}[c]{@{}c@{}} 1.50 \\ (1.23)  \end{tabular}
    & \begin{tabular}[c]{@{}c@{}} 0.29 \\ (0.75)  \end{tabular}
    & \begin{tabular}[c]{@{}c@{}} 9.91 \\ (9.09) \end{tabular}
    & \begin{tabular}[c]{@{}c@{}} 9.98 \\ (9.54) \end{tabular}
    & \begin{tabular}[c]{@{}c@{}} 0.96 \\ (0.97)  \end{tabular}
    \\
    \cline{2-11}
    & \begin{tabular}[c]{@{}c@{}}  LFDPR \\ no AA \end{tabular}
    & \begin{tabular}[c]{@{}c@{}} $\underset{(2.60M)}{6.80}$ \end{tabular}
    & \begin{tabular}[c]{@{}c@{}} 12.17 \end{tabular}
    & \begin{tabular}[c]{@{}c@{}} 0.43 \end{tabular}
    & \bluecell \begin{tabular}[c]{@{}c@{}} $\underset{(1.71\times)}{19.40}$  \end{tabular}
    & \begin{tabular}[c]{@{}c@{}} 2.17 \\ (1.92)  \end{tabular}
    & \begin{tabular}[c]{@{}c@{}} 0.35 \\ (0.74)  \end{tabular}
    & \begin{tabular}[c]{@{}c@{}} 9.70 \\ (8.42) \end{tabular}
    & \begin{tabular}[c]{@{}c@{}} 9.89 \\ (9.07) \end{tabular}
    & \begin{tabular}[c]{@{}c@{}} 0.92 \\ (0.93)  \end{tabular}
    \\
    \cline{2-11}
    & \begin{tabular}[c]{@{}c@{}} LFDPR \\ Sp AA \end{tabular}
    & \begin{tabular}[c]{@{}c@{}} $\underset{(2.60M)}{6.79}$ \end{tabular}
    & \begin{tabular}[c]{@{}c@{}} 12.15 \end{tabular}
    & \begin{tabular}[c]{@{}c@{}} 1.48 \end{tabular}
    & \bluecell \begin{tabular}[c]{@{}c@{}} $\underset{(1.63\times)}{20.42}$ \end{tabular}
    & \begin{tabular}[c]{@{}c@{}} 2.00 \\ (1.87)  \end{tabular}
    & \begin{tabular}[c]{@{}c@{}} 0.35 \\ (0.73)  \end{tabular}
    & \begin{tabular}[c]{@{}c@{}} 9.74 \\ (8.46) \end{tabular}
    & \begin{tabular}[c]{@{}c@{}} 9.91 \\ (9.10) \end{tabular}
    & \begin{tabular}[c]{@{}c@{}} 0.93 \\ (0.93)  \end{tabular}
    \\
    \cline{2-11}
    & \begin{tabular}[c]{@{}c@{}} LFDPR \\ Vi AA \end{tabular}
    & \begin{tabular}[c]{@{}c@{}} $\underset{(2.60M)}{6.81}$ \end{tabular}
    & \begin{tabular}[c]{@{}c@{}} 12.16 \end{tabular}
    & \begin{tabular}[c]{@{}c@{}} 1.25 \end{tabular}
    & \bluecell \begin{tabular}[c]{@{}c@{}} $\underset{(1.64\times)}{20.22}$  \end{tabular}
    & \begin{tabular}[c]{@{}c@{}} 2.18 \\ (1.92)  \end{tabular}
    & \begin{tabular}[c]{@{}c@{}} 0.35 \\ (0.74)  \end{tabular}
    & \begin{tabular}[c]{@{}c@{}} 9.70 \\ (8.42) \end{tabular}
    & \begin{tabular}[c]{@{}c@{}} 9.89 \\ (9.07) \end{tabular}
    & \begin{tabular}[c]{@{}c@{}} 0.92 \\ (0.93)  \end{tabular}
    \\
    \cline{2-11}
    & \begin{tabular}[c]{@{}c@{}} LFDPR \\ Vi-Sp AA \end{tabular}
    & \begin{tabular}[c]{@{}c@{}} $\underset{(2.60M)}{6.79}$ \end{tabular}
    & \begin{tabular}[c]{@{}c@{}} 12.16 \end{tabular}
    & \begin{tabular}[c]{@{}c@{}} 2.14 \end{tabular}
    & \bluecell \begin{tabular}[c]{@{}c@{}} $\underset{(1.58\times)}{21.09}$  \end{tabular}
    & \begin{tabular}[c]{@{}c@{}} 2.05 \\ (1.87)  \end{tabular}
    & \begin{tabular}[c]{@{}c@{}} 0.35 \\ (0.73)  \end{tabular}
    & \begin{tabular}[c]{@{}c@{}} 9.74 \\ (8.46) \end{tabular}
    & \begin{tabular}[c]{@{}c@{}} 9.91 \\ (9.10) \end{tabular}
    & \begin{tabular}[c]{@{}c@{}} 0.93 \\ (0.93)  \end{tabular}
    \\
    \hline
    \hline

    \multicolumn{1}{c|}{\multirow{9}{*}{\shortstack[s]{ Coconut \\ (2.0M) }}} 
    & \begin{tabular}[c]{@{}c@{}}  MVR \end{tabular}
    & \begin{tabular}[c]{@{}c@{}} -  \end{tabular}
    & \begin{tabular}[c]{@{}c@{}} 44.45 \end{tabular}
    & \begin{tabular}[c]{@{}c@{}}  8.23 \end{tabular}
    & \redcell \begin{tabular}[c]{@{}c@{}} 52.68  \end{tabular}
    & \begin{tabular}[c]{@{}c@{}} 0.52 \\ (0.61)  \end{tabular}
    & \begin{tabular}[c]{@{}c@{}} 0.16 \\ (0.75)  \end{tabular}
    & \begin{tabular}[c]{@{}c@{}} 9.98 \\ (9.46) \end{tabular}
    & \begin{tabular}[c]{@{}c@{}} 9.99 \\ (9.76) \end{tabular}
    & \begin{tabular}[c]{@{}c@{}} 0.98 \\ (0.98)  \end{tabular}
    \\
    \cline{2-11}
    & \begin{tabular}[c]{@{}c@{}}  LFDPR \\ no AA \end{tabular}
    & \begin{tabular}[c]{@{}c@{}} $\underset{(1.15M)}{16.53}$ \end{tabular}
    & \begin{tabular}[c]{@{}c@{}} 5.04 \end{tabular}
    & \begin{tabular}[c]{@{}c@{}} 0.41 \end{tabular}
    & \bluecell \begin{tabular}[c]{@{}c@{}} $\underset{(2.40\times)}{21.99}$   \end{tabular}
    & \begin{tabular}[c]{@{}c@{}} 0.59 \\ (0.57)  \end{tabular}
    & \begin{tabular}[c]{@{}c@{}} 0.17 \\ (0.62)  \end{tabular}
    & \begin{tabular}[c]{@{}c@{}} 9.97 \\ (9.54) \end{tabular}
    & \begin{tabular}[c]{@{}c@{}} 9.99 \\ (9.81) \end{tabular}
    & \begin{tabular}[c]{@{}c@{}} 0.98 \\ (0.96)  \end{tabular}
    \\
    \cline{2-11}
    & \begin{tabular}[c]{@{}c@{}} LFDPR \\ Sp AA \end{tabular}
    & \begin{tabular}[c]{@{}c@{}} $\underset{(1.15M)}{16.55}$ \end{tabular}
    & \begin{tabular}[c]{@{}c@{}} 5.02 \end{tabular}
    & \begin{tabular}[c]{@{}c@{}} 1.45 \end{tabular}
    & \bluecell \begin{tabular}[c]{@{}c@{}} $\underset{(2.29\times)}{23.02}$ \end{tabular}
    & \begin{tabular}[c]{@{}c@{}} 0.51 \\ (0.54)  \end{tabular}
    & \begin{tabular}[c]{@{}c@{}} 0.13 \\ (0.56)  \end{tabular}
    & \begin{tabular}[c]{@{}c@{}} 9.99 \\ (9.56) \end{tabular}
    & \begin{tabular}[c]{@{}c@{}} 9.98 \\ (9.83) \end{tabular}
    & \begin{tabular}[c]{@{}c@{}} 0.98 \\ (0.98)  \end{tabular}
    \\
    \cline{2-11}
    & \begin{tabular}[c]{@{}c@{}} LFDPR \\ Vi AA \end{tabular}
    & \begin{tabular}[c]{@{}c@{}} $\underset{(1.51M)}{16.61}$ \end{tabular}
    & \begin{tabular}[c]{@{}c@{}} 5.02 \end{tabular}
    & \begin{tabular}[c]{@{}c@{}} 1.21 \end{tabular}
    & \bluecell \begin{tabular}[c]{@{}c@{}}  $\underset{(2.31\times)}{22.84}$  \end{tabular}
    & \begin{tabular}[c]{@{}c@{}} 0.57 \\ (0.57)  \end{tabular}
    & \begin{tabular}[c]{@{}c@{}} 0.16 \\ (0.62)  \end{tabular}
    & \begin{tabular}[c]{@{}c@{}} 9.97 \\ (9.54) \end{tabular}
    & \begin{tabular}[c]{@{}c@{}} 9.99 \\ (9.81) \end{tabular}
    & \begin{tabular}[c]{@{}c@{}} 0.98 \\ (0.98)  \end{tabular}
    \\
    \cline{2-11}
    & \begin{tabular}[c]{@{}c@{}} LFDPR \\ Vi-Sp AA \end{tabular}
    & \begin{tabular}[c]{@{}c@{}} $\underset{(1.15M)}{16.62}$ \end{tabular}
    & \begin{tabular}[c]{@{}c@{}} 5.02 \end{tabular}
    & \begin{tabular}[c]{@{}c@{}}  2.10 \end{tabular}
    & \bluecell \begin{tabular}[c]{@{}c@{}}  $\underset{(2.22\times)}{23.74}$  \end{tabular}
    & \begin{tabular}[c]{@{}c@{}} 0.50 \\ (0.54)  \end{tabular}
    & \begin{tabular}[c]{@{}c@{}} 0.12 \\ (0.55)  \end{tabular}
    & \begin{tabular}[c]{@{}c@{}} 9.98 \\ (9.59) \end{tabular}
    & \begin{tabular}[c]{@{}c@{}} 9.99 \\ (9.84) \end{tabular}
    & \begin{tabular}[c]{@{}c@{}} 0.98 \\ (0.98)  \end{tabular}
    \\
    \hline
    \hline

    \multicolumn{1}{c|}{\multirow{9}{*}{\shortstack[s]{ Car \\ (300.6K) }}} 
    & \begin{tabular}[c]{@{}c@{}}  MVR \end{tabular}
    & \begin{tabular}[c]{@{}c@{}} -  \end{tabular}
    & \begin{tabular}[c]{@{}c@{}} 38.71 \end{tabular}
    & \begin{tabular}[c]{@{}c@{}} 8.23 \end{tabular}
    & \redcell \begin{tabular}[c]{@{}c@{}} 46.94  \end{tabular}
    & \begin{tabular}[c]{@{}c@{}} 0.21 \\ (0.33)  \end{tabular}
    & \begin{tabular}[c]{@{}c@{}} 0.06 \\ (0.72)  \end{tabular}
    & \begin{tabular}[c]{@{}c@{}} 9.99 \\ (9.73) \end{tabular}
    & \begin{tabular}[c]{@{}c@{}} 9.99 \\ (9.90) \end{tabular}
    & \begin{tabular}[c]{@{}c@{}} 0.99 \\ (0.99)  \end{tabular}
    \\
    \cline{2-11}
    & \begin{tabular}[c]{@{}c@{}}  LFDPR \\ no AA \end{tabular}
    & \begin{tabular}[c]{@{}c@{}} $\underset{(669.4K)}{2.19}$\end{tabular}
    & \begin{tabular}[c]{@{}c@{}} 3.22 \end{tabular}
    & \begin{tabular}[c]{@{}c@{}} 0.42 \end{tabular}
    & \bluecell \begin{tabular}[c]{@{}c@{}} $\underset{(8.06\times)}{5.82}$   \end{tabular}
    & \begin{tabular}[c]{@{}c@{}} 0.52 \\ (0.57)  \end{tabular}
    & \begin{tabular}[c]{@{}c@{}} 0.31 \\ (0.83)  \end{tabular}
    & \begin{tabular}[c]{@{}c@{}} 9.96 \\ (9.53) \end{tabular}
    & \begin{tabular}[c]{@{}c@{}} 9.99 \\ (9.80) \end{tabular}
    & \begin{tabular}[c]{@{}c@{}} 0.98 \\ (0.98)  \end{tabular}
    \\
    \cline{2-11}
    & \begin{tabular}[c]{@{}c@{}} LFDPR \\ Sp AA \end{tabular}
    & \begin{tabular}[c]{@{}c@{}} $\underset{(669.4K)}{2.16}$ \end{tabular}
    & \begin{tabular}[c]{@{}c@{}} 3.21 \end{tabular}
    & \begin{tabular}[c]{@{}c@{}} 1.44\end{tabular}
    & \bluecell \begin{tabular}[c]{@{}c@{}} $\underset{(6.90\times)}{6.81}$   \end{tabular}
    & \begin{tabular}[c]{@{}c@{}} 0.45 \\ (0.54)  \end{tabular}
    & \begin{tabular}[c]{@{}c@{}} 0.32 \\ (0.82)  \end{tabular}
    & \begin{tabular}[c]{@{}c@{}} 9.97 \\ (9.57) \end{tabular}
    & \begin{tabular}[c]{@{}c@{}} 9.99 \\ (9.83) \end{tabular}
    & \begin{tabular}[c]{@{}c@{}} 0.99 \\ (0.99)  \end{tabular}
    \\
    \cline{2-11}
    & \begin{tabular}[c]{@{}c@{}} LFDPR \\ Vi AA \end{tabular}
    & \begin{tabular}[c]{@{}c@{}} $\underset{(669.4K)}{2.17}$ \end{tabular}
    & \begin{tabular}[c]{@{}c@{}} 3.22 \end{tabular}
    & \begin{tabular}[c]{@{}c@{}} 1.21 \end{tabular}
    & \bluecell \begin{tabular}[c]{@{}c@{}} $\underset{(7.12\times)}{6.59}$   \end{tabular}
    & \begin{tabular}[c]{@{}c@{}} 0.50 \\ (0.57)  \end{tabular}
    & \begin{tabular}[c]{@{}c@{}} 0.31 \\ (0.83)  \end{tabular}
    & \begin{tabular}[c]{@{}c@{}} 9.96 \\ (9.53) \end{tabular}
    & \begin{tabular}[c]{@{}c@{}} 9.99 \\ (9.80) \end{tabular}
    & \begin{tabular}[c]{@{}c@{}} 0.98 \\ (0.98)  \end{tabular}
    \\
    \cline{2-11}
    & \begin{tabular}[c]{@{}c@{}} LFDPR \\ Vi-Sp AA \end{tabular}
    & \begin{tabular}[c]{@{}c@{}} $\underset{(669.4K)}{2.20}$ \end{tabular}
    & \begin{tabular}[c]{@{}c@{}} 3.22 \end{tabular}
    & \begin{tabular}[c]{@{}c@{}} 2.09 \end{tabular}
    & \bluecell \begin{tabular}[c]{@{}c@{}} $\underset{(6.25\times)}{7.51}$    \end{tabular}
    & \begin{tabular}[c]{@{}c@{}} 0.44 \\ (0.54)  \end{tabular}
    & \begin{tabular}[c]{@{}c@{}} 0.32 \\ (0.80)  \end{tabular}
    & \begin{tabular}[c]{@{}c@{}} 9.97 \\ (9.58) \end{tabular}
    & \begin{tabular}[c]{@{}c@{}} 9.99 \\ (9.83) \end{tabular}
    & \begin{tabular}[c]{@{}c@{}} 0.99 \\ (0.99)  \end{tabular}
    \\
    \hline
    \hline
    
 \end{tabular}}
\caption{Rendering time and quality results for MVR and LFDPR across four models, as the reconstruction technique (AA or antialiasing) varied between none, spatial, view (angular) and view-spatial. No supersampling was used.}
\label{table:lfd_1xResults}
\end{table}


\begin{landscape}
\setlength{\columnsep}{0.1cm}
\begin{figure}
    \centering
     \includegraphics[width=1.0\linewidth]{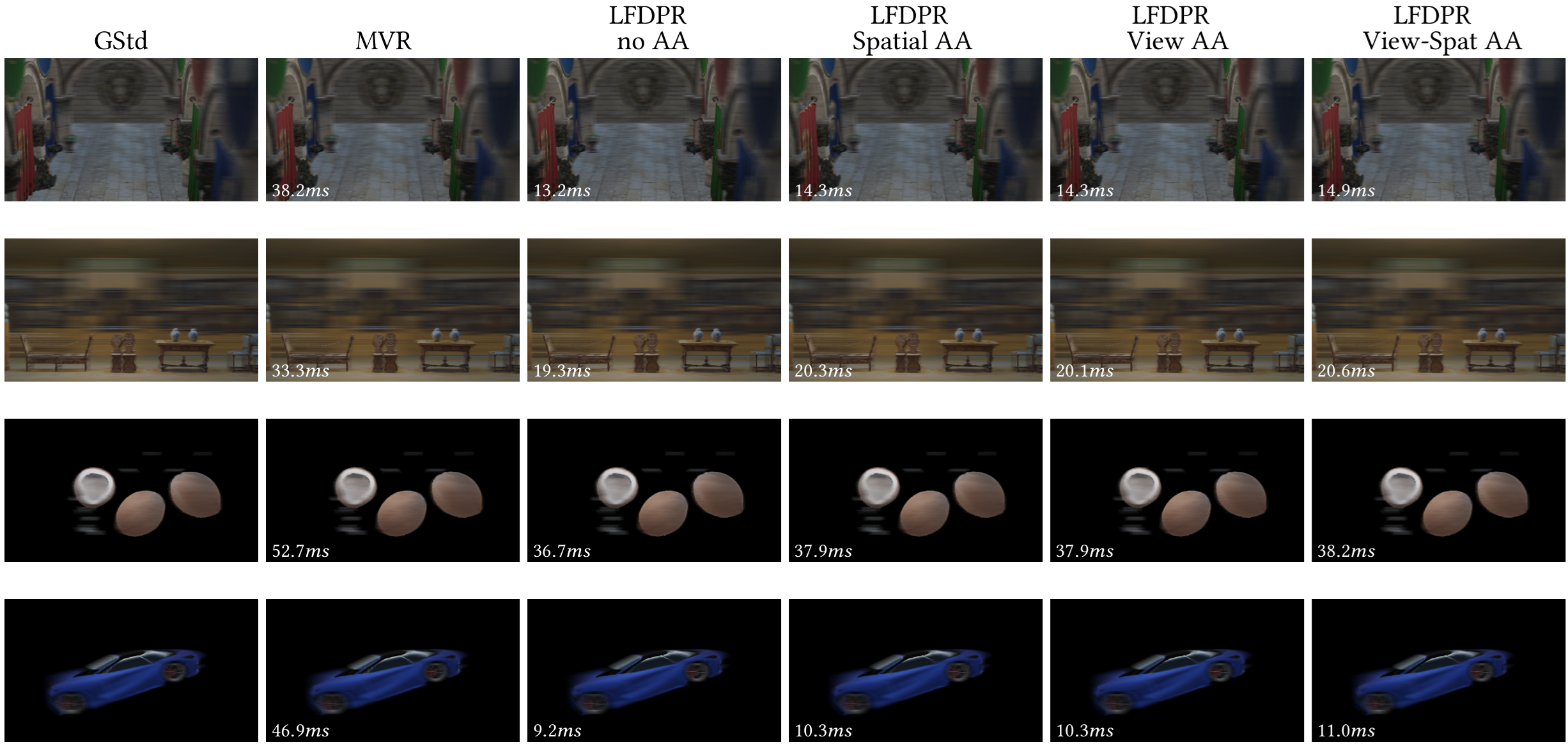}
    \caption{EIA generated using GStd, \ac{mvr}, \ac{lfdpr} without AA, \ac{lfdpr} with spatial AA, \ac{lfdpr} with view AA, and \ac{lfdpr} with view and spatial AA; in the sponza, gallery, coconuts, and car scenes (from top to bottom). All rendering techniques generated 48 views (except GStd with 96 views). \ref{sec:results}.}
    \label{fig:LFD_1xSS}
\end{figure}
\end{landscape}

In \tref{table:lfd_1xResults}, the leftmost and adjacent columns are the name of the scene and the method used to generate the imagery: MVR without anti-aliasing (AA) reconstruction; and LFDPR without AA, with spatial AA, with angular AA, and with spatial-angular AA. Each method generates 48 views with the resolution of 480 $\times$ 360.
The next two columns show \ac{lfdpr}'s point generation time with the number of points in parentheses, and  view generation time.
The fifth and sixth columns report the \ac{eia} construction time and the total rendering time, along with the performance improvement as the ratio of \ac{lfdpr} time over \ac{mvr}’s.
The next five columns present the error and quality measures for both the \ac{eia} and a single view (in parentheses) compared against our gold standard.
The first column of these five reports root mean-squared error (RMSE); the next three report quality measurements from HDR-VDP3 \cite{mantiuk2023hdrvdp3} including probability of detection \textit{P-det}, quality correlate \textit{Q}, and quality correlate in units of just objectionable differences \textit{Q-JOD}; and the last is the structural similarity index measure (SSIM) \cite{wang2004imageSSIM}. 
RMSE ranges from 0 to 255, with lower values better; SSIM scale spans from -1 to 1, with a score of 1 signifing perfect similarity, 0 no similarity, and -1 perfect anti-correlation; \textit{P}-det is probability of detection ranging from 0 to 1, with lower values better;  \textit{Q} (quality correlate) ranges from 10 at best quality down to negative values; and lastly \textit{Q-JOD} is similar to \textit{Q} but scaled to JOD units, each corresponding to $75$\% of the population noticing the difference between the pair of images \cite{perez2019pairwiseQ-JOD}. 

\ac{lfdpr} renders these views up to $8\times$ faster than \ac{mvr} with comparable image quality. In the sponza scene, \ac{lfdpr}'s \textit{P-det} was thrice as good as \ac{mvr}'s. LFDPR speed slowed as the number of triangles grew (see the coconuts), particularly when the number of texels per pixel was high (coconuts and gallery). Our mid-range GPU has limited bandwidth, which began saturating when the number of points was particularly high (gallery). LFDPR's quality was always close to and sometimes exceeded MVR's quality.

\tref{table:lfd_2xResults} shows the results when using $2\times$ supersampling in space (48 views, each with a resolution of 960 x 720) and views (96 views, each with a resolution of 480 x 360). To reach the displayed resolution, MVR performs a pixel-by-pixel unweighted average of the supersampled image. The leftmost column names the scene, and the adjacent two indicate either spatial or view supersampling, and the rendering method. The remaining columns follow \tref{table:lfd_1xResults}'s layout. Note that in MVR, for each view we generate mipmap, resulting in multiple texture bindings during the EIA construction pass and causing longer latency on memory buses, consequently slowing speed for EIA construction.

Here, LFDPR is up to $3\times$ faster for spatial supersampling, and up to $5\times$ faster for view supersampling, while maintaining nearly equivalent quality. In one case (spatial supersampling with the sponza), LFDPR became slightly slower than MVR. Note that spatial supersampling slows LFDPR more than angular, since filling two times more pixels can require roughly two times more points, while projecting the same set of points into random nearby views did not increase point cloud size. The reduction in rendering speed to achieve supersampling was not matched with a proportional improvement in quality. In general, spatial supersampling improved quality more than angular supersampling. However, angular supersampling is an essentially multiview technique, while our quality measures are single view techniques. We hope to assess the impact of angular supersampling more accurately using studies with human observers.

\subsection{Quality Comparison}

As we noted, while speed varied greatly, quality did not. \fref{fig:LFD_1xSS} shows the differences in the sponza (top row), gallery (second), coconuts (third) and car (bottom row) as rendered by GStd (left column), MVR (second), LFDPR (third), LFDPR with spatial AA (fourth), with view AA (fifth), and view-spatial AA (rightmost column).  The visual quality of \ac{lfdpr} is quite comparable to \ac{mvr} in all of these and LFDPR renders them much more quickly.


\begin{table}[t]

\centering
\resizebox{\linewidth}{!}{
 \begin{tabular}{ c | c | c | c  c  c  c | c  c  c  c  c|}
 \hline
     \multicolumn{1}{c|}{\multirow{3}{*}{\shortstack[s]{Scene \\ (\#tris)}}} 
   & \multicolumn{1}{c|}{\multirow{3}{*}{\shortstack[s]{SS}}} 
   & \multicolumn{1}{c|}{\multirow{3}{*}{\shortstack[s]{ Method }}} 
   & \multicolumn{4}{c|}{ Performance (in milliseconds)    }
   & \multicolumn{5}{c|}{ Error / Quality for EIA (Single View) }
   \\
   \cline{4-12} 

    & 
    & 
    & \begin{tabular}[c]{@{}c@{}} pt gen \\ \textsuperscript{(\#pts)}  \end{tabular}
    & \begin{tabular}[c]{@{}c@{}} view gen \end{tabular}
    & \begin{tabular}[c]{@{}c@{}} EIA \\ constr \end{tabular}
    & \begin{tabular}[c]{@{}c@{}} tot \\ time  \end{tabular}
    & \begin{tabular}[c]{@{}c@{}} RMSE  \end{tabular}
    & \begin{tabular}[c]{@{}c@{}} P-det  \end{tabular}
    & \begin{tabular}[c]{@{}c@{}} Q  \end{tabular}
    & \begin{tabular}[c]{@{}c@{}} Q-JOD  \end{tabular}
    & \begin{tabular}[c]{@{}c@{}} SSIM  \end{tabular}
    \\ 
    \hline \hline

    \multicolumn{1}{c|}{\multirow{8}{*}{\shortstack[s]{ Sponza \\ (1.0M) }}} 
    & \multicolumn{1}{c|}{\multirow{4}{*}{\shortstack[s]{ Spatial }}} 
    & \begin{tabular}[c]{@{}c@{}} MVR \end{tabular}
    & \begin{tabular}[c]{@{}c@{}} -  \end{tabular}
    & \begin{tabular}[c]{@{}c@{}} 33.33 \end{tabular}
    & \begin{tabular}[c]{@{}c@{}} 8.21 \end{tabular}
    & \bluecell \begin{tabular}[c]{@{}c@{}} 41.54  \end{tabular}
    & \begin{tabular}[c]{@{}c@{}} 1.43 \\ (0.98)  \end{tabular}
    & \begin{tabular}[c]{@{}c@{}} 0.34 \\ (0.29)  \end{tabular}
    & \begin{tabular}[c]{@{}c@{}} 9.92 \\ (9.44) \end{tabular}
    & \begin{tabular}[c]{@{}c@{}} 9.98 \\ (9.75) \end{tabular}
    & \begin{tabular}[c]{@{}c@{}} 0.94 \\ (0.97)  \end{tabular}
    \\
    \cline{3-12}
    &
    & \begin{tabular}[c]{@{}c@{}} LFDPR \\ Sp AA \end{tabular}
    & \begin{tabular}[c]{@{}c@{}} $\underset{(6.66M)}{5.14}$ \end{tabular}
    & \begin{tabular}[c]{@{}c@{}} 29.54 \end{tabular}
    & \begin{tabular}[c]{@{}c@{}} 7.42 \end{tabular}
    & \redcell \begin{tabular}[c]{@{}c@{}} $\underset{(0.99\times)}{42.10}$  \end{tabular}
    & \begin{tabular}[c]{@{}c@{}} 1.68 \\ (1.52)  \end{tabular}
    & \begin{tabular}[c]{@{}c@{}} 0.29 \\ (0.48)  \end{tabular}
    & \begin{tabular}[c]{@{}c@{}} 9.89 \\ (8.93) \end{tabular}
    & \begin{tabular}[c]{@{}c@{}} 9.97 \\ (9.43) \end{tabular}
    & \begin{tabular}[c]{@{}c@{}} 0.92 \\ (0.93)  \end{tabular}
    \\
    \cline{2-12}
    & \multicolumn{1}{c|}{\multirow{4}{*}{\shortstack[s]{ View  }}} 
    & \begin{tabular}[c]{@{}c@{}} MVR \end{tabular}
    & \begin{tabular}[c]{@{}c@{}} -  \end{tabular}
    & \begin{tabular}[c]{@{}c@{}} 69.04 \end{tabular}
    & \begin{tabular}[c]{@{}c@{}} 11.56 \end{tabular}
    & \redcell \begin{tabular}[c]{@{}c@{}} 80.60  \end{tabular}
    & \begin{tabular}[c]{@{}c@{}} 1.83 \\ (1.52)  \end{tabular}
    & \begin{tabular}[c]{@{}c@{}} 0.74 \\ (0.73)  \end{tabular}
    & \begin{tabular}[c]{@{}c@{}} 9.84 \\ (8.82) \end{tabular}
    & \begin{tabular}[c]{@{}c@{}} 9.95 \\ (9.36) \end{tabular}
    & \begin{tabular}[c]{@{}c@{}} 0.91 \\ (0.93)  \end{tabular}
    \\
    \cline{3-12}
    &
    & \begin{tabular}[c]{@{}c@{}} LFDPR \\ Vi AA \end{tabular}
    & \begin{tabular}[c]{@{}c@{}} $\underset{(2.37M)}{2.89}$ \end{tabular}
    & \begin{tabular}[c]{@{}c@{}} 21.33 \end{tabular}
    & \begin{tabular}[c]{@{}c@{}} 7.13 \end{tabular}
    & \bluecell \begin{tabular}[c]{@{}c@{}} $\underset{(2.57\times)}{31.34}$  \end{tabular}
    & \begin{tabular}[c]{@{}c@{}} 2.10 \\ (1.90)  \end{tabular}
    & \begin{tabular}[c]{@{}c@{}} 0.32 \\ (0.67)  \end{tabular}
    & \begin{tabular}[c]{@{}c@{}} 9.75 \\ (8.58) \end{tabular}
    & \begin{tabular}[c]{@{}c@{}} 9.91 \\ (9.19) \end{tabular}
    & \begin{tabular}[c]{@{}c@{}} 0.88 \\ (0.90)  \end{tabular}
    \\
    \cline{3-12}
    \hline 
    \hline

    \multicolumn{1}{c|}{\multirow{8}{*}{\shortstack[s]{ Gallery \\ (998.9K) }}} 
    & \multicolumn{1}{c|}{\multirow{4}{*}{\shortstack[s]{ Spatial }}} 
    & \begin{tabular}[c]{@{}c@{}} MVR \end{tabular}
    & \begin{tabular}[c]{@{}c@{}} -  \end{tabular}
    & \begin{tabular}[c]{@{}c@{}} 31.08 \end{tabular}
    & \begin{tabular}[c]{@{}c@{}} 8.22 \end{tabular}
    & \redcell \begin{tabular}[c]{@{}c@{}} 39.30  \end{tabular}
    & \begin{tabular}[c]{@{}c@{}} 1.21 \\ (0.89)  \end{tabular}
    & \begin{tabular}[c]{@{}c@{}} 0.08 \\ (0.54)  \end{tabular}
    & \begin{tabular}[c]{@{}c@{}} 9.96 \\ (9.42) \end{tabular}
    & \begin{tabular}[c]{@{}c@{}} 9.99 \\ (9.74) \end{tabular}
    & \begin{tabular}[c]{@{}c@{}} 0.97 \\ (0.98)  \end{tabular}
    \\
    \cline{3-12}
    &
    & \begin{tabular}[c]{@{}c@{}} LFDPR \\ Sp AA \end{tabular}
    & \begin{tabular}[c]{@{}c@{}} $\underset{(5.57M)}{7.35}$ \end{tabular}
    & \begin{tabular}[c]{@{}c@{}} 22.26 \end{tabular}
    & \begin{tabular}[c]{@{}c@{}} 7.26 \end{tabular}
    & \bluecell \begin{tabular}[c]{@{}c@{}} $\underset{(1.07\times)}{36.86}$ \end{tabular}
    & \begin{tabular}[c]{@{}c@{}} 1.40 \\ (1.32)  \end{tabular}
    & \begin{tabular}[c]{@{}c@{}} 0.13 \\ (0.54)  \end{tabular}
    & \begin{tabular}[c]{@{}c@{}} 9.94 \\ (8.99) \end{tabular}
    & \begin{tabular}[c]{@{}c@{}} 9.99 \\ (9.47) \end{tabular}
    & \begin{tabular}[c]{@{}c@{}} 0.96 \\ (0.96)  \end{tabular}
    \\
    \cline{2-12}
    & \multicolumn{1}{c|}{\multirow{4}{*}{\shortstack[s]{ View  }}} 
    & \begin{tabular}[c]{@{}c@{}} MVR \end{tabular}
    & \begin{tabular}[c]{@{}c@{}} -  \end{tabular}
    & \begin{tabular}[c]{@{}c@{}} 49.51 \end{tabular}
    & \begin{tabular}[c]{@{}c@{}} 12.39 \end{tabular}
    & \redcell \begin{tabular}[c]{@{}c@{}} 61.90  \end{tabular}
    & \begin{tabular}[c]{@{}c@{}} 1.38 \\ (1.23)  \end{tabular}
    & \begin{tabular}[c]{@{}c@{}} 0.29 \\ (0.75)  \end{tabular}
    & \begin{tabular}[c]{@{}c@{}} 9.91 \\ (9.09) \end{tabular}
    & \begin{tabular}[c]{@{}c@{}} 9.98 \\ (9.54) \end{tabular}
    & \begin{tabular}[c]{@{}c@{}} 0.96 \\ (0.97)  \end{tabular}
    \\
    \cline{3-12}
    &
    & \begin{tabular}[c]{@{}c@{}} LFDPR \\ Vi AA \end{tabular}
    & \begin{tabular}[c]{@{}c@{}} $\underset{(2.60M)}{7.01}$ \end{tabular}
    & \begin{tabular}[c]{@{}c@{}} 23.78 \end{tabular}
    & \begin{tabular}[c]{@{}c@{}} 7.14 \end{tabular}
    & \bluecell \begin{tabular}[c]{@{}c@{}} $\underset{(1.63\times)}{37.93}$ \end{tabular}
    & \begin{tabular}[c]{@{}c@{}} 2.01 \\ (1.92)  \end{tabular}
    & \begin{tabular}[c]{@{}c@{}} 0.34 \\ (0.66)  \end{tabular}
    & \begin{tabular}[c]{@{}c@{}} 9.75 \\ (8.39) \end{tabular}
    & \begin{tabular}[c]{@{}c@{}} 9.91 \\ (9.04) \end{tabular}
    & \begin{tabular}[c]{@{}c@{}} 0.93 \\ (0.93)  \end{tabular}
    \\
    \cline{3-12}
    \hline 
    \hline

    \multicolumn{1}{c|}{\multirow{8}{*}{\shortstack[s]{ Coconuts \\ (2.0M) }}} 
    & \multicolumn{1}{c|}{\multirow{4}{*}{\shortstack[s]{ Spatial }}} 
    & \begin{tabular}[c]{@{}c@{}} MVR \end{tabular}
    & \begin{tabular}[c]{@{}c@{}} -  \end{tabular}
    & \begin{tabular}[c]{@{}c@{}} 50.41 \end{tabular}
    & \begin{tabular}[c]{@{}c@{}} 8.26 \end{tabular}
    & \redcell \begin{tabular}[c]{@{}c@{}} 58.67 \end{tabular}
    & \begin{tabular}[c]{@{}c@{}} 0.37 \\ (0.31)  \end{tabular}
    & \begin{tabular}[c]{@{}c@{}} 0.06 \\ (0.37)  \end{tabular}
    & \begin{tabular}[c]{@{}c@{}} 9.99 \\ (9.85) \end{tabular}
    & \begin{tabular}[c]{@{}c@{}} 9.99 \\ (9.95) \end{tabular}
    & \begin{tabular}[c]{@{}c@{}} 0.99 \\ (0.99)  \end{tabular}
    \\
    \cline{3-12}
    &
    & \begin{tabular}[c]{@{}c@{}} LFDPR \\ Sp AA \end{tabular}
    & \begin{tabular}[c]{@{}c@{}} $\underset{(1.16M)}{17.05}$ \end{tabular}
    & \begin{tabular}[c]{@{}c@{}} 5.27 \end{tabular}
    & \begin{tabular}[c]{@{}c@{}} 6.98 \end{tabular}
    & \bluecell \begin{tabular}[c]{@{}c@{}} $\underset{(2.0\times)}{29.30}$  \end{tabular}
    & \begin{tabular}[c]{@{}c@{}} 0.40 \\ (0.42)  \end{tabular}
    & \begin{tabular}[c]{@{}c@{}} 0.04 \\ (0.34)  \end{tabular}
    & \begin{tabular}[c]{@{}c@{}} 9.99 \\ (9.75) \end{tabular}
    & \begin{tabular}[c]{@{}c@{}} 9.99 \\ (9.91) \end{tabular}
    & \begin{tabular}[c]{@{}c@{}} 0.99 \\ (0.99)  \end{tabular}
    \\
    \cline{2-12}
    & \multicolumn{1}{c|}{\multirow{4}{*}{\shortstack[s]{ View  }}} 
    & \begin{tabular}[c]{@{}c@{}} MVR \end{tabular}
    & \begin{tabular}[c]{@{}c@{}} -  \end{tabular}
    & \begin{tabular}[c]{@{}c@{}} 89.07 \end{tabular}
    & \begin{tabular}[c]{@{}c@{}} 10.83 \end{tabular}
    & \redcell \begin{tabular}[c]{@{}c@{}} 99.91 \end{tabular}
    & \begin{tabular}[c]{@{}c@{}} 0.50 \\ (0.61)  \end{tabular}
    & \begin{tabular}[c]{@{}c@{}} 0.14 \\ (0.75)  \end{tabular}
    & \begin{tabular}[c]{@{}c@{}} 9.98 \\ (9.46) \end{tabular}
    & \begin{tabular}[c]{@{}c@{}} 9.99 \\ (9.76) \end{tabular}
    & \begin{tabular}[c]{@{}c@{}} 0.98 \\ (0.98)  \end{tabular}
    \\
    \cline{3-12}
    &
    & \begin{tabular}[c]{@{}c@{}} LFDPR \\ Vi AA \end{tabular}
    & \begin{tabular}[c]{@{}c@{}}  $\underset{(1.14M)}{16.88}$  \end{tabular}
    & \begin{tabular}[c]{@{}c@{}} 10.22 \end{tabular}
    & \begin{tabular}[c]{@{}c@{}} 6.85 \end{tabular}
    & \bluecell \begin{tabular}[c]{@{}c@{}} $\underset{(2.94\times)}{33.95}$ \end{tabular}
    & \begin{tabular}[c]{@{}c@{}} 0.51 \\ (0.55)  \end{tabular}
    & \begin{tabular}[c]{@{}c@{}} 0.12 \\ (0.68)  \end{tabular}
    & \begin{tabular}[c]{@{}c@{}} 9.98 \\ (9.58) \end{tabular}
    & \begin{tabular}[c]{@{}c@{}} 9.99 \\ (9.83) \end{tabular}
    & \begin{tabular}[c]{@{}c@{}} 0.98 \\ (0.98)  \end{tabular}
    \\
    \cline{3-12}
    \hline 
    \hline

    \multicolumn{1}{c|}{\multirow{8}{*}{\shortstack[s]{ Car \\ (300.6K) }}} 
    & \multicolumn{1}{c|}{\multirow{4}{*}{\shortstack[s]{ Spatial }}} 
    & \begin{tabular}[c]{@{}c@{}} MVR \end{tabular}
    & \begin{tabular}[c]{@{}c@{}} -  \end{tabular}
    & \begin{tabular}[c]{@{}c@{}} 46.22 \end{tabular}
    & \begin{tabular}[c]{@{}c@{}} 8.21 \end{tabular}
    & \redcell \begin{tabular}[c]{@{}c@{}} 54.44 \end{tabular}
    & \begin{tabular}[c]{@{}c@{}} 0.21 \\ (0.19)  \end{tabular}
    & \begin{tabular}[c]{@{}c@{}} 0.04 \\ (0.33)  \end{tabular}
    & \begin{tabular}[c]{@{}c@{}} 9.99 \\ (9.90) \end{tabular}
    & \begin{tabular}[c]{@{}c@{}} 9.99 \\ (9.97) \end{tabular}
    & \begin{tabular}[c]{@{}c@{}} 0.99 \\ (0.99)  \end{tabular}
    \\
    \cline{3-12}
    &
    & \begin{tabular}[c]{@{}c@{}} LFDPR \\ Sp AA \end{tabular}
    & \begin{tabular}[c]{@{}c@{}} $\underset{(1.82M)}{2.60}$ \end{tabular}
    & \begin{tabular}[c]{@{}c@{}} 7.92 \end{tabular}
    & \begin{tabular}[c]{@{}c@{}} 7.24 \end{tabular}
    & \bluecell \begin{tabular}[c]{@{}c@{}} $\underset{(3.07\times)}{17.76}$ \end{tabular}
    & \begin{tabular}[c]{@{}c@{}} 0.32 \\ (0.36)  \end{tabular}
    & \begin{tabular}[c]{@{}c@{}} 0.05 \\ (0.48)  \end{tabular}
    & \begin{tabular}[c]{@{}c@{}} 9.99 \\ (9.80) \end{tabular}
    & \begin{tabular}[c]{@{}c@{}} 9.99 \\ (9.93) \end{tabular}
    & \begin{tabular}[c]{@{}c@{}} 0.99 \\ (0.99)  \end{tabular}
    \\
    \cline{2-12}
    & \multicolumn{1}{c|}{\multirow{4}{*}{\shortstack[s]{ View  }}} 
    & \begin{tabular}[c]{@{}c@{}} MVR \end{tabular}
    & \begin{tabular}[c]{@{}c@{}} -  \end{tabular}
    & \begin{tabular}[c]{@{}c@{}} 75.27 \end{tabular}
    & \begin{tabular}[c]{@{}c@{}} 10.92 \end{tabular}
    & \redcell \begin{tabular}[c]{@{}c@{}} 86.19 \end{tabular}
    & \begin{tabular}[c]{@{}c@{}} 0.20 \\ (0.33)  \end{tabular}
    & \begin{tabular}[c]{@{}c@{}} 0.05 \\ (0.72)  \end{tabular}
    & \begin{tabular}[c]{@{}c@{}} 9.99 \\ (9.73) \end{tabular}
    & \begin{tabular}[c]{@{}c@{}} 9.99 \\ (9.90) \end{tabular}
    & \begin{tabular}[c]{@{}c@{}} 0.99 \\ (0.99)  \end{tabular}
    \\
    \cline{3-12}
    &
    & \begin{tabular}[c]{@{}c@{}} LFDPR \\ Vi AA \end{tabular}
    & \begin{tabular}[c]{@{}c@{}} $\underset{(669.4K)}{2.22}$ \end{tabular}
    & \begin{tabular}[c]{@{}c@{}} 7.02 \end{tabular}
    & \begin{tabular}[c]{@{}c@{}} 7.21 \end{tabular}
    & \bluecell \begin{tabular}[c]{@{}c@{}} $\underset{(5.24\times)}{16.45}$  \end{tabular}
    & \begin{tabular}[c]{@{}c@{}} 0.45 \\ (0.54)  \end{tabular}
    & \begin{tabular}[c]{@{}c@{}} 0.32 \\ (0.81)  \end{tabular}
    & \begin{tabular}[c]{@{}c@{}} 9.97 \\ (9.56) \end{tabular}
    & \begin{tabular}[c]{@{}c@{}} 9.99 \\ (9.82) \end{tabular}
    & \begin{tabular}[c]{@{}c@{}} 0.99 \\ (0.99)  \end{tabular}
    \\
    \cline{3-12}
    \hline 
    \hline

 \end{tabular}}
\caption{Results structured similarly to \tref{table:lfd_1xResults}, but with all renderers using $2\times$ supersampling. LFDPR did not use combined spatial-angular AA here.}
\vspace{-10pt}
\label{table:lfd_2xResults}
\end{table}

\begin{landscape}
\setlength{\columnsep}{0.1cm}
\begin{figure}[h]
\centering
 \includegraphics[width=1.0\linewidth]{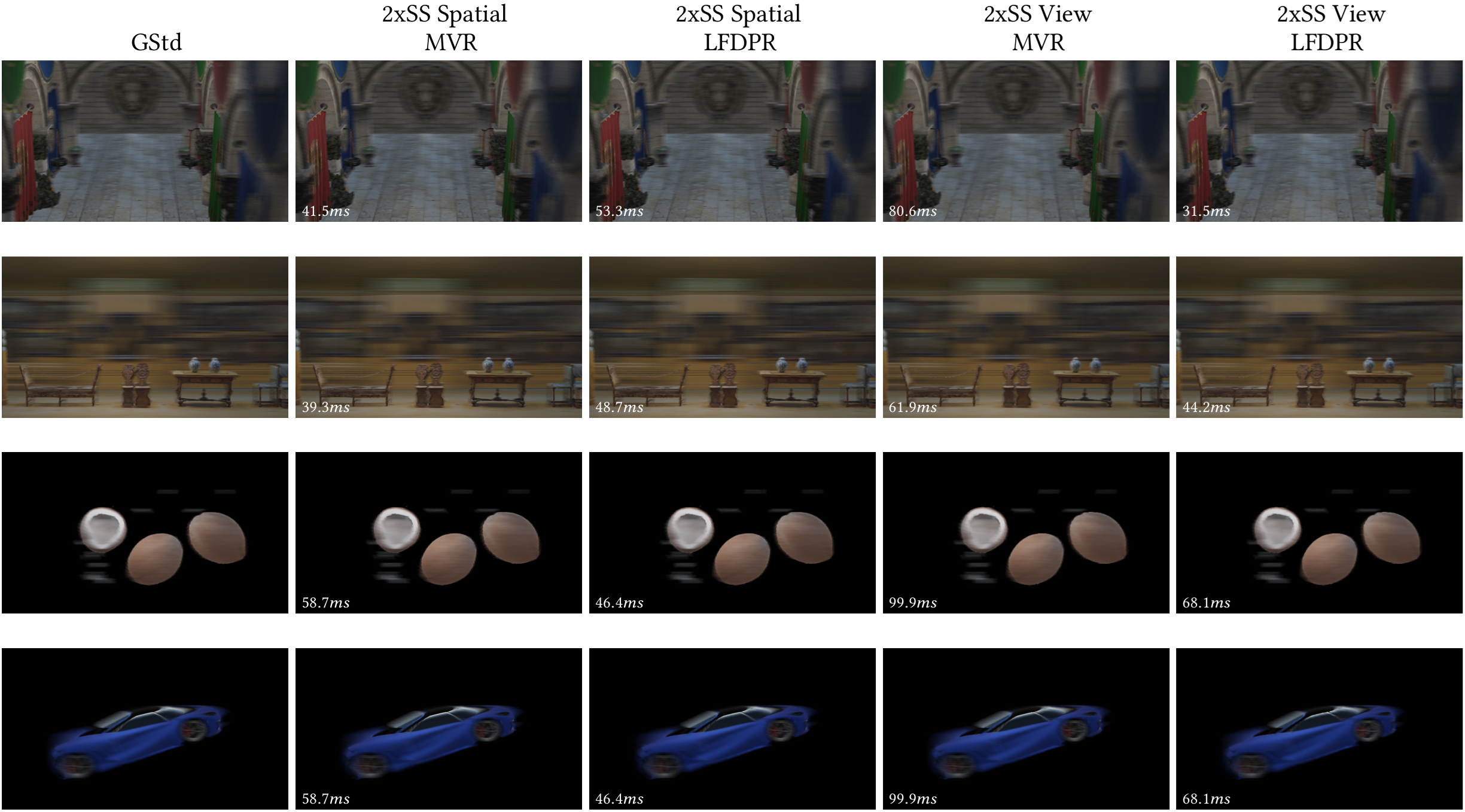}
\caption{EIA generated using GStd, $2\times$ spatial SS \ac{mvr} and \ac{lfdpr} with spatial AA, $2\times$ view SS \ac{mvr} and \ac{lfdpr} with view AA; in the sponza, gallery, coconuts, and car scenes (from top to bottom). $2\times$ spatial SS generates 48 views at a resolution of $960\times720$, and $2\times$ view SS generates $96$ views at a resolution of $480\times360$.}
\label{fig:LFD_2xSS}
\end{figure}

\end{landscape}

\begin{figure}[ht]
    \centering
     \includegraphics[width=1.0\linewidth]{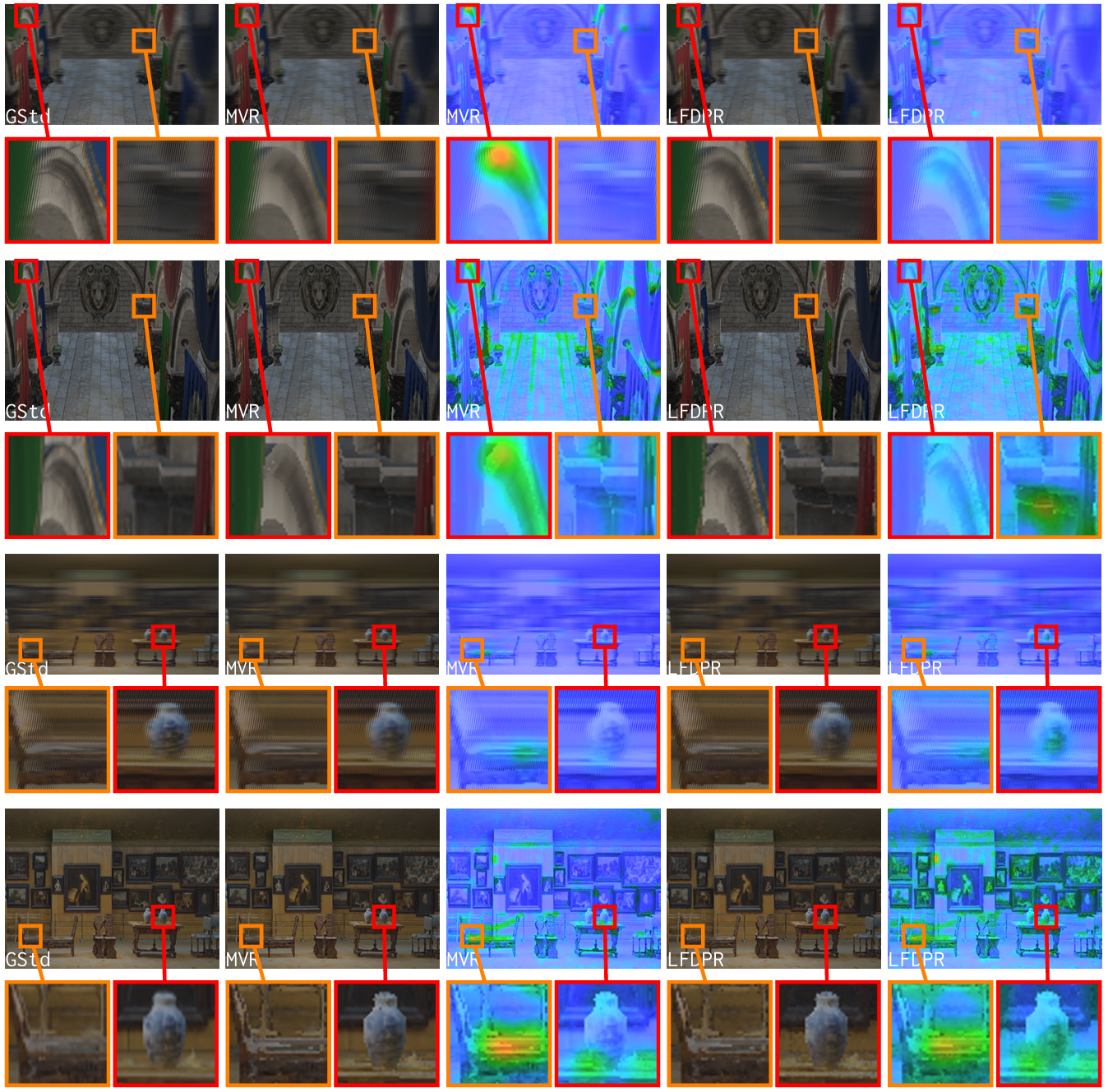}  
    \caption{ Quality comparison of GStd vs MVR and LFDPR with angular-spatial reconstruction. The leftmost column shows the GStd, the adjacent and central column show MVR and its HDR-VDP3 diifference image, and the right two columns show LFDPR and its difference image. The top four rows are the sponza, the bottom two the gallery. For both the scenes, the first tow rows shows EIA generated image and the next two rows shows one of the single view (SV).}
    \vspace{-10pt}
    \label{fig:LFD_1xSS_Quality}      
\end{figure}

\begin{figure}[ht]
    \centering
         \includegraphics[width=1.0\linewidth]{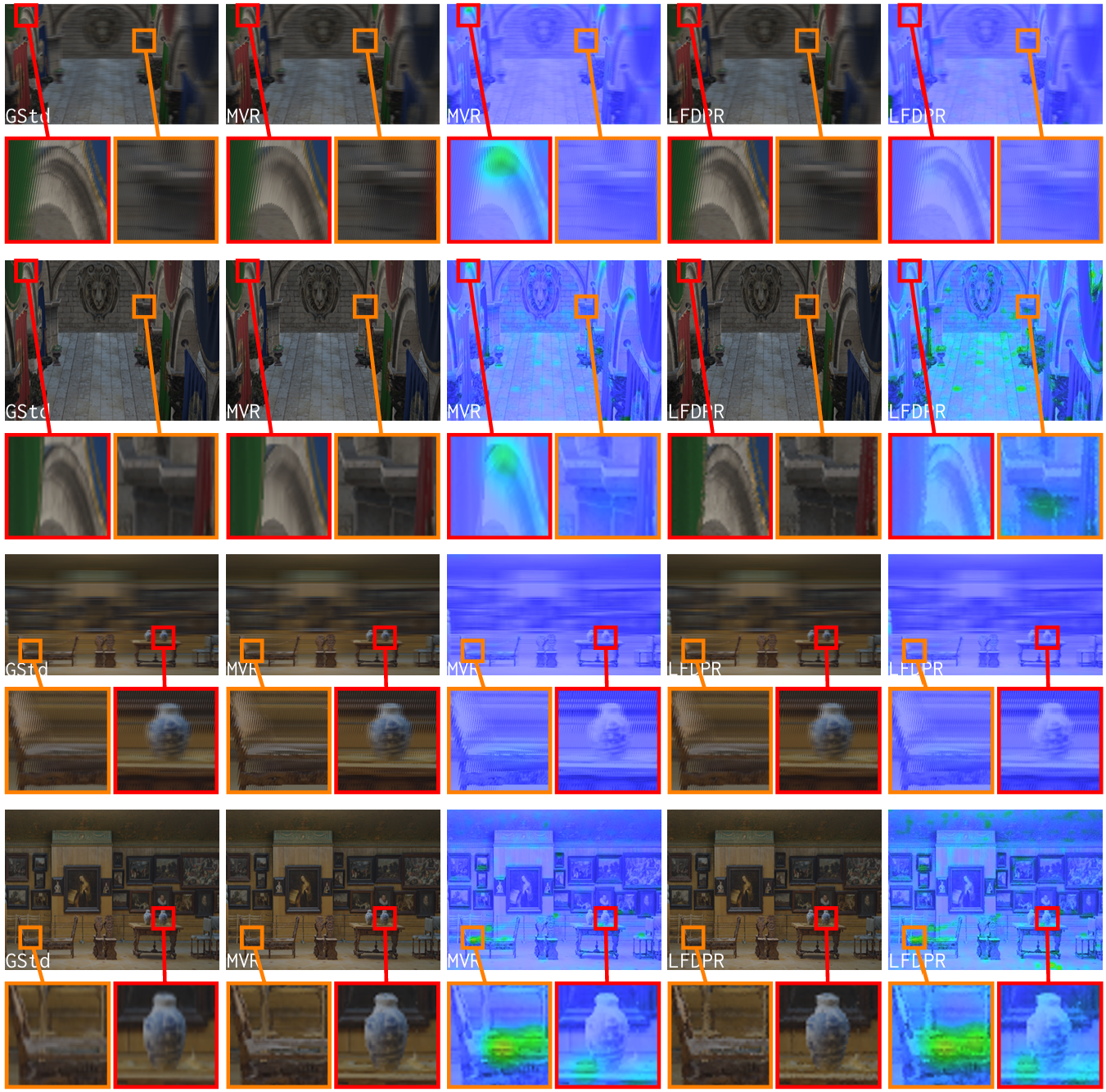}  
    \caption{ Quality comparison of GStd vs MVR and LFDPR with angular-spatial reconstruction, both with 2x spatial supersampling. The layout is the same as in \fref{fig:LFD_1xSS_Quality}.}
    \vspace{-10pt}
    \label{fig:LFD_2xSS_Quality}      
\end{figure}

\fref{fig:LFD_1xSS_Quality} zooms in on quality differences between MVR and LFDPR with angular-spatial reconstruction. Here, GStd is in the leftmost column, MVR and its HDR-VDP3 difference are in the second and third columns, and LFDPR is in the fourth and rightmost columns. LFDPR renders fine details in the sponza (top four rows), more clearly than MVR does. This can be seen clearly in the second row's red boxes. This may be due to LFDPR's flexible multiview mipmapping.

\fref{fig:LFD_2xSS} is laid out similarly to \fref{fig:LFD_1xSS}, but uses $2\times$ supersampling. These images look nearly identical to those in \fref{fig:LFD_1xSS}, despite being rendered much more slowly. Zooming in on the differences in \fref{fig:LFD_2xSS_Quality} does not change this impression much, though the error images do reveal a reduction in error salience. Again, while supersampling did bring a modest improvement in quality, it came at a steep price in rendering speed. This tradeoff might be mitigated on a GPU with higher bandwidth buses.

        \section{Conclusions and Future Work}
\label{sec:conclusion_futurework} 
Rendering for light field displays (LFDs) requires creating dozens or hundreds of views, which must then be combined into a single EIA image on the display. Multiview rendering at such scales is extremely challenging, making real-time LFD rendering extremely difficult.

Our innovations meet these challenges by marshaling and improving on the techniques of both iVIR and EPR.
Unlike many multiview effects, all of the pixels generated for LFD rendering are (potentially) visible, making the eye-based efficiencies of EPR insignificant. We address these problems by introducing texture-based splatting, which avoids oversampling of triangles mapped to only a few texels; and with LFD-biased sampling, which adjusts horizontal and vertical triangle sampling to match the sampling of the LFD itself.
LFD rendering's "every pixel" visibility not only makes high frame rates more difficult, it also makes higher quality imagery more difficult. To produce such quality, we introduce multiview mipmapping, which reduces texture aliasing even though compute shaders do not support hardware mipmapping; and use EPR's splatting, which avoids iVIR's edge bloat \cite{iVIR2022}. We also introduce angular supersampling and reconstruction to combat LFD view aliasing and crosstalk.
The resulting LFDPR is 2-8$\times$ times faster than MVR, with similar comparable quality. Texture splatting and multiview mipmapping may also prove to be valuable in EPR.

In future work, we hope to examine LFDPR speed on higher bandwidth GPUs, which should make rendering with spatial supersampling much more performant. We also plan to study the effect on quality of angular reconstruction and supersampling in human experiments, since the error measures and single frame comparisons we present here do not capture the effects of view change well. In particular, we hope that these view angle quality improvements may reduce crosstalk, an LFD limitation in which LFD users see two images at the same time.


\begin{acks}
We would like to extend our heartfelt gratitude to Dr. Hee-Jin Choi for generously providing the LFD prototype display crucial for our research.
This material is based upon work supported by the National Science Foundation under Grant No. - 2008590.
\end{acks}
   
    \bibliographystyle{ACM-Reference-Format} 

\end{document}